\documentclass[aps,pra,twocolumn,superscriptaddress,nofootinbib]{revtex4-2}

\usepackage{amsmath,amssymb,mathtools}
\usepackage{bbm}
\usepackage{amsthm}
\usepackage{booktabs}
\usepackage{array}
\usepackage{graphicx}
\usepackage[colorlinks=true,linkcolor=blue,citecolor=blue,urlcolor=blue]{hyperref}
\usepackage{xurl}
\usepackage{orcidlink}

\newtheorem{definition}{Definition}

\newtheorem{remark}{Remark}

\newcommand{\C}{\mathcal{C}}
\newcommand{\E}{\mathcal{E}}

\newcommand{\HH}{\mathcal{H}}
\newcommand{\OmegaSet}{\Omega}
\newcommand{\TVS}{\mathcal{S}_2}
\newcommand{\Ptwo}{\mathsf{P}_{2}}

\newcommand{\Cone}{\operatorname{Cone}}
\newcommand{\conv}{\operatorname{conv}}

\newcommand{\tr}{\operatorname{tr}}
\newcommand{\id}{\mathbbm{1}}
\newcommand{\reals}{\mathbb{R}}
\newcommand{\complex}{\mathbb{C}}

\newcommand{\YO}{\mathrm{YO}}

\begin{document}

\title{Which Classicality? Incidence, Simplex, and Product-Rule Tests in Finite Quantum Logics}

\author{Karl Svozil \orcidlink{0000-0001-6554-2802}}
\affiliation{Institute for Theoretical Physics, TU Wien,
Wiedner Hauptstra{\ss}e 8-10/136, 1040 Vienna, Austria}
\email{karl.svozil@tuwien.ac.at}

\date{\today}

\begin{abstract}
Finite quantum-logical constructions can appear classical or nonclassical
depending on which structure is retained.  We distinguish incidence tests based
on valuations, colorings, and partition representations; simplex-embedding
tests for specified prepare-and-measure fragments; and operator-functional
tests imposing spectral and product rules.  We show that the collective GHZ
joint measurement is a single Boolean context and becomes nonclassical only
when a common assignment of local factors, together with product preservation,
is required.  For selected labelled ray fragments, we compute
state-depolarizing thresholds for a restricted projector-cone factorization
and for a specified vector-generated operational closure, separating exact
primal--dual certificates from numerical estimates.  These values are
properties of the stated fragments and noise model, not invariants of the
underlying hypergraphs or a universal ordering of contextuality.
\end{abstract}

\maketitle

\section{Introduction}

In a single maximal measurement context there is no immediate operational
separation between the classical and quantum descriptions of one block.
Classically, such a block is a partition, equivalently an equivalence relation
on a finite set; quantum mechanically it is represented by a Hilbert-space
projection-valued measure (PVM).  A Boolean block \(C=\{P_1,\ldots,P_d\}\)
produces detector-click frequencies \(p_i\geq0\) with \(\sum_i p_i=1\).  The
same one-context statistics can be read classically as a probability measure on
the atoms of a partition, or quantum mechanically as Born probabilities
\(p_i=\langle\psi|P_i|\psi\rangle\) for a pure state and a rank-one orthogonal
context.

Both classical and quantum state spaces are closed under probabilistic mixing:
if \(P_1\) and \(P_2\) are allowed preparations, then
\(tP_1+(1-t)P_2\) is an allowed preparation for \(0\leq t\leq1\).  Moreover,
the probability of every outcome is the corresponding convex mixture,
\[
        p(e|tP_1+(1-t)P_2)
        =
        t\,p(e|P_1)+(1-t)\,p(e|P_2).
\]
In this precise sense both classical and quantum probabilities are affine
under probabilistic mixing.  Their state-space geometries nevertheless
differ: a classical normalized state space is a simplex, whereas a quantum
state space is not, and pure-state transition probabilities are squared
projection lengths, for example of Malus type.

For a fixed preparation, even a continuous response curve such as
\(p(+|\theta)=\cos^2\theta\) can be reproduced by a setting-by-setting
classical description: on a common ontic space one may stipulate a
response function \(\xi_{+|\theta}(\lambda)\) whose average has that value~\cite{aerts-69}.
By itself, therefore, the curvature of a one-parameter response is not a
simplex obstruction.  However, such a construction leaves unrestricted how the
responses for different settings, and the ontic distributions for different
preparations, are related.  A nonclassical obstruction arises only when one
requires a single model to satisfy additional cross-context constraints:
for example, common preparation representations for operationally equivalent
mixtures, common response representations for operationally equivalent
effects, transformation-composition relations, or Bell locality, where the
same local response functions and one shared hidden-variable distribution must
work for all choices of local settings.  Bell violations rule out precisely
such a global Bell-local factorization, not a setting-by-setting fit to an
isolated response curve~\cite{svozil-2026-os}.

The same distinction explains why two contexts, even when pasted along one or
more common (intertwining) atoms, do not by themselves produce an operational obstruction.
Let two \(m\)-atom Boolean blocks share a proper subset \(S\) of atoms,
\(0<|S|<m\).  If the two probability assignments agree on every shared atom,
then the total probability left for the nonshared atoms is the same in both
blocks.  That remaining mass can always be coupled by an ordinary joint
distribution over predetermined answers for the two measurements.  Thus bare
two-context statistics, including Firefly-type pastings along a single atom,
still have the same operational shadow as a classical marginal model.

A nonclassical probability-level obstruction appears only when a larger family
of local distributions is required to be the family of marginals of one global
nonnegative distribution, or when a prepare-and-measure table is required to
factor through one classical simplex, and that extension or factorization
problem becomes infeasible.

Farkas' lemma~\cite{Farkas1902,Schrijver}, in the marginal-extension form
emphasized by Garg and Mermin~\cite{Garg1984}, concerns the question of
whether specified lower-order or context-wise distributions admit compatible
higher-order distributions with those marginals.  Farkas' alternative says
that either such a nonnegative extension exists, or there is a separating
linear inequality certifying that no extension exists.  Bell~\cite{bell},
Clauser--Horne (CH)~\cite{cl-horne}, and
Klyachko--Can--Binicio\u{g}lu--Shumovsky (KCBS)~\cite{Klyachko-2008}
inequalities are instances of this general marginal-extension logic; the
simplex-embeddability linear program (LP) used below is its prepare-and-measure
version.

The preceding local observations should therefore not be interpreted as a
classical reconstruction of quantum theory.  They identify where, and under
which limited circumstances, the one-context and (intertwining) two-context shadows may
be considered classically reproducible, and where additional operational or
marginal constraints must enter.

Once a sufficiently rich family of contexts
is retained, the familiar Hilbert-space obstructions reappear.  In Hilbert
spaces of dimension at least three, quantum
propositions cannot in general be assigned pre-existing, context-independent
truth values compatible with the functional structure of sharp
measurements~\cite{Gleason,specker-60,kamber65,ZirlSchl-65,Zierler1975,
kochen1,cabello-96,Pavi_i__2019,cabello2021contextuality}.  In quantum-logical
language this is first a statement about pasted Boolean algebras, orthomodular
posets, or orthogonality hypergraphs.  At the operational level it becomes a
question about whether a specified prepare-and-measure table admits a classical
explanation~\cite{froissart-81,Garg1984,pitowsky-86,svozil-2001-cesena,
Klyachko-2008,Yu-2012,peres222}.
Generalized-noncontextual ontological models~\cite{Spekkens-04}, the
simplex-embeddability criterion of Schmid, Selby, Wolfe, Kunjwal, and
Spekkens~\cite{schmid-2021-simplex}, and the LP of Selby, Wolfe,
Schmid, Sainz, and Rossi~\cite{selby-2024-lp} give one precise way to formulate
that operational question.  Operator-valued parity proofs, such as
Greenberger--Horne--Zeilinger (GHZ) and Mermin arguments~\cite{ghz,mermin90b,
svozil-2024-convert-pra-externalfigures}, retain still another part of the same
Hilbert-space description: algebraic relations among observables.

The aim of this paper is to keep these retained structures visible.  Incidence data lead to valuation, coloring, and partition-logic questions~\cite{kochen1,svozil-2001-eua,svozil-2025-color}.  Convex-operational data lead to simplex-embedding questions for specified preparations, effects, probabilities, and operational equivalences.  Operator data lead to functional-calculus and product-rule questions for assigned observable values.  These are related tests, but they are not a hierarchy.  The same physical construction can change its classicality status when contexts, effects, local factors, coarse grainings, or operational equivalences are added or forgotten.
Our central finding is that classicality is not a binary property of a quantum logical structure, but a bookkeeping choice: which data (incidence, convex-operational, or operator-algebraic) are retained determines which test applies and what obstruction appears.

Five consequences of this bookkeeping guide the paper.  A hypergraph can be
classically representable as a partition logic while a quantum realization of
the same pasting supports nonclassical Born probabilities.  A closed fragment
consisting of one Boolean context is always simplex-classical.  GHZ is not a
contradiction within one Boolean joint spectral context; it appears when global
product outcomes are identified with products of pre-existing local Pauli
values.  Simplex embeddability is not the same question as KS uncolorability,
and chromatic contextuality is a distinct combinatorial obstruction from
ordinary scarcity of two-valued states.

This organization complements existing classifications rather than replacing them.  The graph-theoretic approach of Cabello, Severini, and Winter relates exclusivity graphs to invariants such as independence numbers and the Lov\'asz theta function~\cite{cabello-severini-winter-2014}.  The hypergraph approach of Ac\'in, Fritz, Leverrier, and Sainz formulates contextuality scenarios directly as compatibility/exclusivity structures with probabilistic models~\cite{acin-fritz-leverrier-sainz-2015}.  The sheaf-theoretic framework of Abramsky and Brandenburger classifies empirical models by the obstruction to gluing local sections, distinguishing probabilistic, logical, and strong forms of contextuality~\cite{abramsky-brandenburger-2011}.  The present distinction is cross-cutting: before applying a classicality test, it asks whether one has retained only incidence data, an operational probability fragment, or operator-algebraic product data.

The numerical computations below therefore use two related fragments.  First, a deliberately restricted projector-cone proxy uses the listed rank-one projectors as both preparations and effects, with state-side depolarizing noise.  Second, for the rational vector sets, a vector-generated operational closure adds unit effects, residual outcomes of PVMs, complements, coarse grainings, and an optional preparation closure by normalized coarse-grained events~\cite{schmid-2021-simplex,selby-2024-lp}.  The first audit gives exact-certified benchmarks for several projector cones.  The second shows how the numbers change when some of the operational data omitted by the proxy are restored.

The detailed examples are deferred to the body of the paper.  GHZ and Peres--Mermin are discussed where operator products and masked contexts are analyzed; meager two-valued-state spaces and chromatic colorability are treated in their own combinatorial sections; and the table of thresholds is interpreted only after the computational protocol has been specified.  The introduction therefore only fixes the bookkeeping: which data are being kept, and which classicality test is then being applied.

The paper proceeds as follows.  Section~\ref{sec:three-layers} defines the three retained structures.  Section~\ref{sec:simplex-test} recalls the linear-programming test for simplex embeddability, and Section~\ref{sec:sandwich} gives its general orthant-factorization interpretation.  Section~\ref{sec:operator-valued} explains how operator-valued simplex tests depend on whether operators are treated as spectral-outcome data or as algebraic quantities subject to product rules.  Section~\ref{sec:two-contexts} explains why two contexts, or weakly pasted contexts, need not by themselves separate vector probabilities from classical convex probabilities.  The remaining sections discuss calibration examples, meager two-valued-state spaces, chromatic colorability, the computational protocol, the projector-cone thresholds, and the vector-generated closure comparison.

\section{Three retained structures and their tests}
\label{sec:three-layers}

\subsection{Combinatorial contextuality}

Combinatorial contextuality is a property of the pasted context structure.  Its input is an orthogonality hypergraph $H=(V,\C)$ whose vertices are atoms and whose hyperedges are contexts.  The basic question is whether there exist global context-independent assignments compatible with the local Boolean structure.  The most familiar assignments are two-valued states
\begin{equation}
        s:V\to\{0,1\},
        \qquad \sum_{v\in C}s(v)=1\quad(C\in\C).
\end{equation}
Their absence gives the Kochen--Specker (KS) obstruction.  Their scarcity gives the meager cases considered below: nonunital, nonseparating, and nonfull two-valued-state spaces.  The term \emph{meager} is used here only as a convenient umbrella term for these finite-state-space pathologies; it is not meant in the Baire-category sense.  Their abundance, especially a separating family, permits a concrete set representation and hence a partition-logic model~\cite{svozil-2001-eua}.

Chromatic contextuality~\cite{svozil-2025-color} also belongs to the combinatorial level, but it asks for more than the existence of valuations.  A strong $d$-coloring of a $d$-uniform hypergraph decomposes the constant-one assignment into $d$ two-valued states, one per color.  Thus chromatic contextuality is a spectral-labeling obstruction: there is no way to use one fixed set of $d$ nondegenerate outcome labels globally across all contexts.  This is a discrete global obstruction, not a probability-table obstruction.

Partition logics---families of partitions, equivalently families of
equivalence relations, on a finite set with generalized urn or automaton
realizations~\cite{svozil-2001-eua,svozil-2018-b}---sit at the opposite
end of the same combinatorial level.  Their existence demonstrates that
non-Boolean pasting and complementarity can also be represented on an
underlying classical point space.  Each partition is the set of equivalence
classes of an equivalence relation on the underlying set, and a partition
logic is obtained by taking a family of such partitions and identifying
equal blocks across different partitions.  Its dispersion-free states are
the point evaluations determined by the underlying elements of that set.
In Wright-style generalized urn models, the same construction is realized
by ball types and color filters; in Moore or Mealy automata, it is realized
by initial states and input-dependent output partitions.  Such models
implement complementarity without value indefiniteness~\cite{schaller-92,svozil-2018-b}.

\subsection{Operational simplex contextuality}

Simplex contextuality is an operational convex-geometric notion.  Its input is not just a hypergraph but a finite prepare-and-measure fragment
\begin{equation}
        (\OmegaSet,\E,p(e|\rho)),
\end{equation}
where $\OmegaSet$ is a set of preparations, $\E$ is a set of effects, and $p(e|\rho)$ is the observed probability table.  In the quantum cases below, $p(e|\rho)=\tr(e\rho)$ and the preparations and effects are rank-one projectors associated with the listed rays.

A finite table by itself is not the nontrivial object: without operational
equivalences one can always simulate it by taking one ontic state per
preparation and copying the observed response probabilities.  The nontrivial
question is whether one positive model can extend linearly to the generated
preparation and effect cones while respecting the operational equivalences that
are part of the fragment.

A fragment is simplex embeddable when its specified pairing factors through a
classical simplex.  When the unit effect, normalization, and all relevant
preparation- and measurement-event equivalences are included, this is
equivalent to a generalized-noncontextual model of that operational fragment.
For a restricted projector-cone calculation that omits some of those data, it
is more precise to call the result a positive cone-factorization benchmark.
In either case the property depends on which preparations and effects are
included.  Enlarging a ray set can change the generated cones and hence the
reported robustness, even if the added projectors are an orthogonal completion
of an independently specified hypergraph.

The incidence and operational tests can coincide, but need not.  A KS set such
as the Cabello 18--9 configuration is nonclassical at the valuation level and,
for the natural projector fragment, simplex contextual as well.  A partition
logic, however, can be fully classical at the incidence level while a quantum
realization with the same pasting has different Born probabilities.

Conversely, a restricted operational fragment may fail to reveal a defect
already present in the underlying hypergraph.  Nonunitality is exposed only if
the offending atom can actually be prepared or measured; nonseparability only
if included preparations statistically distinguish the identified atoms; and
a true-implies-false (TIFS) or nonfullness constraint only if the relevant
endpoint preparations and effects are present.  Without those operational
witnesses, the simplex test may remain feasible even though the incidence
structure has a combinatorial defect.

\subsection{Operator-functional contextuality}

Operator-functional contextuality, as the term is used here, is not a new rule beyond Kochen--Specker functional composition.  It is a separation of that familiar requirement into two parts.  The observables are treated as algebraic quantities, not merely as names for their spectral projectors.

The first is \emph{functional-calculus preservation} (FUNC):
\begin{equation}
        v(f(A))=f(v(A)).
        \label{eq:func-rule}
\end{equation}
For a sharp observable with spectral resolution
\begin{equation}
        A=\sum_i a_iP_i,
\end{equation}
this condition is largely equivalent to choosing exactly one value-one spectral projector in the resolved Boolean algebra generated by the $P_i$.  When the relevant spectral projectors and their Boolean relations are already part of the combinatorial input, Eq.~\eqref{eq:func-rule} is therefore mostly reflected in the two-valued-state framework.

The second requirement is \emph{multiplicative preservation} for commuting observables:
\begin{equation}
        v(AB)=v(A)v(B),
        \qquad [A,B]=0.
        \label{eq:product-rule}
\end{equation}
The product rule is not independent of FUNC when the whole commuting algebra is included: \(AB\) is then a function of the joint spectral observable.  The operational point here is narrower.  After maximal spectral refinement of one context, ordinary projector valuations capture only the Boolean structure of that context, whereas GHZ and Peres--Mermin arguments also identify operator factors across different compatible products.  This is the genuinely operator-algebraic ingredient in those parity arguments.  It relates different operator presentations that share factors, and it is not captured by looking only at the joint spectral projectors of a single resolved context.  Thus the third layer is best understood as having two subcomponents: spectral-functional consistency, which often collapses to projector-valued logic after maximal refinement, and multiplicative consistency, which can impose additional parity constraints across commuting operator products.

This distinction is decisive for GHZ-type arguments.  The Mermin--GHZ operators
\begin{equation}
X_1X_2X_3,\quad X_1Y_2Y_3,\quad Y_1X_2Y_3,\quad Y_1Y_2X_3
\end{equation}
commute and therefore possess a common eigenbasis.  That eigensystem is one orthonormal basis in $\complex^8$; as a projector logic it is a single Boolean block with eight atoms and therefore has eight dispersion-free states~\cite{svozil-2024-convert-pra-externalfigures}.  There is no KS contradiction at the level of that isolated context.  The contradiction enters only when the value of each global product observable is assumed to factor into values of the local Pauli components $X_i$ and $Y_i$.  Then each local factor appears twice in the product of the four assigned values, forcing $+1$ classically, whereas the corresponding operator product is $-\id$.  The failure is therefore not the absence of a valuation on one context, but the impossibility of a multiplicative, product-preserving valuation on the operator algebra being implicitly invoked.
The classical contradiction arises only when one assumes that the value of a global product observable equals the product of pre-existing local values, an assumption that holds in noncontextual ontological models but is operationally unjustified for incompatible local settings.

The Peres--Mermin square~\cite{peres111,mermin90b,peres-91}
is different but related.  At the operator level it is a parity proof.
When the masked contexts are extracted by simultaneous diagonalization or by matrix pencils,
the operator argument can be transcribed into a projective KS structure,
in particular its associated 24--24 (vector--context) completion~\cite{Pavicic-2005,svozil-2024-convert-pra-externalfigures}  containing the Cabello--Estebaranz--Garc\'ia-Alcaine 18--9 set~\cite{cabello-96,svozil-2024-convert-pra-externalfigures}.  Thus some multiplicative operator contradictions descend to combinatorial projector contextuality after unmasking, while others, such as the single-context GHZ eigensystem, do not.  This is why the operator-functional layer deserves to be kept separate from both colorability and simplex embeddability, while still recognizing that its functional-calculus part overlaps with the projector-valued combinatorial layer after full spectral resolution.

The word ``single-context'' is therefore a statement about one particular operational packaging, not about the full GHZ reasoning.  The collective measurement of the four global products is a single Boolean context.  The local GHZ experiment, however, invokes four incompatible local settings and the counterfactual identification of local factors across them.  For the collective packaging the simplex and two-valued-state tests are trivial; for the local-factor packaging the nonclassicality is absolute and rests on locality, noncontextuality, and product preservation.  Equivalently, the collective product observable and its local-factor implementation should not be identified without further assumptions.  A collective measurement reports a global spectral outcome; a local-factor implementation refines that outcome into local Pauli outcomes and then imposes a product rule on their assigned values.

\subsection{Probability theories on the same pasting}

The probability level is itself not exhausted by the binary alternative ``classical or quantum.''  On a fixed exclusivity structure one may consider at least three types of context-independent additive weights.  First, classical probabilities are convex mixtures of dispersion-free states; they form a polytope.  Second, Born probabilities arise from a faithful orthogonal representation and a state vector or density operator; graph-theoretically, these are related to Lov\'asz-type theta-body constructions.  Third, there are exotic weights, such as Wright-type weights on pentagon-like logics~\cite{greechie-1974,wright:pent}, which satisfy additivity within contexts but are neither classical convex-hull points nor Hilbert-space Born probabilities.

Thus the incidence structure alone does not determine the probability calculus.  The same pasted logic may admit a partition model, a vector representation, and additional exotic additive weights.  This is the sense in which combinatorial contextuality and simplex contextuality are largely independent diagnostics.  The combinatorial test classifies global assignments on the logic; the simplex test concerns a particular operational probability model.

\section{The operational simplex test}
\label{sec:simplex-test}

Let $\OmegaSet=\{\rho_1,\ldots,\rho_N\}$ be a finite set of quantum
preparations, and let $\E=\{e_1,\ldots,e_M\}$ be a finite set of quantum
effects.  Here each $\rho_i$ is a positive semidefinite preparation operator
on $\HH$, and each $e_j$ is an effect, $0\le e_j\le I$.  The word
``state'' is therefore used in the operational quantum sense, not in the
sense of a two-valued state or valuation on an orthogonality hypergraph.

The preparations may be normalized or subnormalized.  Here subnormalized
means
\begin{equation}
        \rho_i\ge 0,
        \qquad
        0\le \tr(\rho_i)\le 1.
\end{equation}
Thus one may write $\rho_i=p_i\widehat{\rho}_i$, where
$\widehat{\rho}_i$ is a normalized density operator and
$0\le p_i\le 1$ is a weight or success probability.  This convention is
natural for cone constructions, since the cones below are closed under
nonnegative rescaling.

Let \(\HH\) be the finite-dimensional Hilbert space of the fragment, and let
\begin{equation}
        \mathcal A=\operatorname{Herm}(\HH)
\end{equation}
denote the real vector space of Hermitian operators on $\HH$.  We do not
assume that the chosen coordinates on $\mathcal A$ are trace-orthonormal.
Instead, choose an arbitrary real basis
\begin{equation}
        B_1,\ldots,B_q
\end{equation}
of $\mathcal A$.  If
\begin{equation}
        A=\sum_{\alpha=1}^q a_\alpha B_\alpha,
        \qquad
        C=\sum_{\beta=1}^q c_\beta B_\beta,
\end{equation}
then the trace pairing is represented by the Gram matrix
\begin{equation}
        G_{\alpha\beta}=\tr(B_\alpha B_\beta).
        \label{eq:trace-gram-matrix}
\end{equation}
Thus, in these coordinates,
\begin{equation}
        \tr(AC)=a^T G c.
        \label{eq:trace-pairing-coordinate}
\end{equation}
Only in a trace-orthonormal basis would $G$ be the identity matrix.

The following construction concerns a finite prepare-and-measure
fragment: no preparation equivalences beyond normalization, no Bell-locality
constraint, and no additional contexts or cross-context relations are imposed.

Let
\begin{equation}
        S_\Omega=\operatorname{span}_{\mathbb R}(\OmegaSet),
        \qquad
        S_E=\operatorname{span}_{\mathbb R}(\E)
\end{equation}
be the accessible preparation and effect subspaces of the real vector space
$\mathcal A$ of Hermitian operators on $\HH$.  Choose arbitrary real bases of
$S_\Omega$ and $S_E$, and let
\begin{equation}
        d_\Omega=\dim S_\Omega,
        \qquad
        d_E=\dim S_E.
\end{equation}

For the factorization statements below, the restricted Born pairing is
understood operationally.  Thus either it is separating on the chosen
accessible spaces, or one first passes to the quotients
\begin{equation}
        \overline S_\Omega=
        S_\Omega/(S_\Omega\cap S_E^\perp),
        \qquad
        \overline S_E=
        S_E/(S_E\cap S_\Omega^\perp),
        \label{eq:operational-quotients}
\end{equation}
where orthogonality is with respect to the trace pairing.  These quotients
identify preparations or effects that the other side of the fragment cannot
distinguish.  If a quotient is nontrivial, choose linear sections
\begin{equation}
        j_\Omega:\overline S_\Omega\longrightarrow S_\Omega\subseteq\mathcal A,
        \qquad
        j_E:\overline S_E\longrightarrow S_E\subseteq\mathcal A
\end{equation}
of the quotient maps.  If the pairing is already separating, these maps are
just the canonical inclusions.  Thus the choice of representatives is explicit
when it is needed; the induced Born pairing on the quotient spaces does not
depend on that choice.  To avoid notational clutter, bars are suppressed below,
and the dimensions and bases are understood after this reduction.  A numerical
implementation should verify the separating condition, or perform the indicated
quotient, before interpreting the result as an operational simplex test.

After the choices of bases and sections, let
\begin{equation}
        I_\Omega\in\mathbb R^{q\times d_\Omega},
        \qquad
        I_E\in\mathbb R^{q\times d_E}.
\end{equation}
Concretely, the columns of $I_\Omega$ and $I_E$ are the ambient coordinates,
in the basis $B_1,\ldots,B_q$ of $\mathcal A$, of the selected
representatives of the reduced preparation and effect bases.

Therefore, if $x\in\reals^{d_\Omega}$ is the coordinate vector of an
operational preparation class and $y\in\reals^{d_E}$ is that of an operational
effect class, then $I_\Omega x$ and $I_E y$ are chosen ambient
representatives.  The Born pairing between their classes is
\begin{equation}
        \tr(e_y\rho_x)
        =
        (I_E y)^T G (I_\Omega x)
        =
        y^T I_E^T G I_\Omega x.
        \label{eq:born-pairing-coordinate}
\end{equation}
Thus
\begin{equation}
        B_{E\Omega}=I_E^T G I_\Omega
        \label{eq:born-pairing-matrix}
\end{equation}
is the matrix of the accessible state--effect pairing in the chosen reduced
coordinates.

For the original finite fragment, this pairing gives the operational table
\begin{equation}
        p_{ij}=\tr(e_j\rho_i),
        \qquad
        1\le i\le N,\quad 1\le j\le M .
\end{equation}
For normalized preparations these entries are probabilities; for
subnormalized preparations they are the corresponding probability weights.
Thus the pair $(\OmegaSet,\E)$ determines only a finite operational image, or
its convex hull, in the corresponding probability space.

Note that a Lov\'asz-theta-type body
~\cite{GroetschelLovaszSchrijver1986,Cabello-2014-gtatqc} is not determined
by the finite preparation set $\OmegaSet$ alone.  It is associated with an
exclusivity structure together with an orthogonal representation, or, in the
present projective setting, with a fixed family of projective effects carrying
the relevant orthogonality relations.  Varying the density operator over all
quantum states gives the corresponding quantum probability body.  Restricting
the density operator to the finite set $\OmegaSet$ produces only a finite
operational image, or its convex hull, inside that body. The theta-type body describes the quantum probability region associated with
an exclusivity structure and an orthogonal representation, whereas the simplex
test asks whether a specified finite part of that region factors through a
classical simplex.  Thus simplex nonembeddability means that the chosen
operational fragment lies outside every classical simplex image compatible
with the specified preparation and effect cones.

After choosing the reduced bases of $S_\Omega$ and $S_E$, let $V_\Omega$ and
$V_E$ denote the matrices whose columns are the coordinate vectors of the
$\rho_i$ and $e_j$, respectively.  The state and effect cones are first given
by their generator, or $V$-, representations:
\begin{align}
        \Cone[\OmegaSet]
        &=
        \left\{
        V_\Omega r : r\in\reals^N,\ r_i\ge 0
        \right\}                                                   \nonumber\\
        &=
        \left\{
        \sum_{i=1}^N r_i\rho_i : r_i\ge 0
        \right\}
        \subseteq S_\Omega,
        \label{eq:state-cone-vrep}
        \\
        \Cone[\E]
        &=
        \left\{
        V_E s : s\in\reals^M,\ s_j\ge 0
        \right\}                                                   \nonumber\\
        &=
        \left\{
        \sum_{j=1}^M s_j e_j : s_j\ge 0
        \right\}
        \subseteq S_E.
        \label{eq:effect-cone-vrep}
\end{align}

Because these cones are finitely generated, they are polyhedral.
By the Minkowski--Weyl theorem, every finitely generated polyhedral cone also
admits a finite half-space, or \hbox{$H$-,} representation, and conversely~\cite{ziegler,Henk-Ziegler-polytopes,Avis:1997:GCH:280651.280652,mcmullen-71,Schrijver,gruenbaum-2003,Fukuda-techrep}.
This stage is similar to the standard computation of Bell-type inequalities, like the Clauser--Horne--Shimony--Holt inequalities~\cite{froissart-81,pitowsky-86}.
Those $H$-representations are linear
functionals
\begin{equation}
        h^\Omega_1,\ldots,h^\Omega_{n_\Omega}\in S_\Omega^*,
        \qquad
        h^E_1,\ldots,h^E_{n_E}\in S_E^*,
\end{equation}
which may be chosen as facet-defining inequalities, such that
\begin{align}
        \Cone[\OmegaSet]
        &=
        \left\{
        x\in S_\Omega :
        h^\Omega_k(x)\ge 0
        \ \text{for } k=1,\ldots,n_\Omega
        \right\},
        \label{eq:state-cone-hrep}
        \\
        \Cone[\E]
        &=
        \left\{
        y\in S_E :
        h^E_\ell(y)\ge 0
        \ \text{for } \ell=1,\ldots,n_E
        \right\}.
        \label{eq:effect-cone-hrep}
\end{align}
Equivalently, collect these functionals as the rows of linear maps
\begin{equation}
        H_\Omega:S_\Omega\to\reals^{n_\Omega},
        \qquad
        H_E:S_E\to\reals^{n_E},
\end{equation}
so that
\begin{equation}
        H_\Omega x
        =
        \begin{pmatrix}
        h^\Omega_1(x)\\
        \vdots\\
        h^\Omega_{n_\Omega}(x)
        \end{pmatrix},
        \qquad
        H_E y
        =
        \begin{pmatrix}
        h^E_1(y)\\
        \vdots\\
        h^E_{n_E}(y)
        \end{pmatrix}.
\end{equation}
With this notation,
\begin{align}
        H_\Omega x\ge_{\mathrm e}0
        &\Longleftrightarrow
        x\in\Cone[\OmegaSet],
        \nonumber\\
        H_E y\ge_{\mathrm e}0
        &\Longleftrightarrow
        y\in\Cone[\E].
        \label{eq:facet-map-criterion}
\end{align}
Here $\ge_{\mathrm e}$ denotes entrywise nonnegativity: all components of the
corresponding vector must be nonnegative.  It is not the positive-semidefinite
order on Hermitian operators.

The simplex-embeddability linear program asks whether the Born-pairing matrix
\eqref{eq:born-pairing-matrix} factors through the facet descriptions of the
state and effect cones.  Equivalently, one asks whether there exists an
entrywise nonnegative matrix
\begin{equation}
        \sigma\in\reals^{n_E\times n_\Omega}
\end{equation}
such that
\begin{equation}
        I_E^T G I_\Omega
        =
        H_E^T\sigma H_\Omega,
        \qquad
        \sigma\ge_{\mathrm e}0.
        \label{eq:simplex-lp}
\end{equation}
The dimensions are
\begin{equation}
        I_E^T G I_\Omega\in\reals^{d_E\times d_\Omega},
        \qquad
        H_E^T\sigma H_\Omega\in\reals^{d_E\times d_\Omega}.
\end{equation}
Thus both sides represent the same bilinear pairing between accessible effects
and accessible states.

This is the usual simplex embedding written in cone language.  The map
\(x\mapsto H_\Omega x\) sends accessible preparations positively into an
orthant, and every positive linear map from a polyhedral cone to an orthant is
generated by nonnegative combinations of the facet functionals.  Dually,
\(y\mapsto H_E y\) supplies the positive response functionals.  The
nonnegative matrix \(\sigma\) then couples the two orthants so that the
resulting bilinear form agrees with the Born pairing on the accessible
quotient spaces.

When Eq.~\eqref{eq:simplex-lp} is feasible, the specified pairing admits a
positive factorization through a finite orthant.  With the unit effect,
normalization constraints, operational equivalences, and relevant
coarse-grainings included, this is a generalized-noncontextual simplex
embedding of the corresponding operational fragment.  Without those data it
is more accurately described as a positive cone-factorization benchmark.

In particular, if the projector-cone proxy omits \(u-e\) for an effect \(e\),
the LP enforces positivity of the response to \(e\), but not necessarily the
operational upper bound \(\xi_e\leq1\).  The projector-cone thresholds below
are therefore not full operational noncontextuality robustnesses.  They become
instances of the full Schmid--Selby--Wolfe--Kunjwal--Spekkens test only when the
operational scenario supplies the unit effect, complements, coarse-grainings,
normalization, and the preparation- and measurement-event equivalences at
issue.

Equivalently, the cone factorization becomes an ordinary normalized simplex
ontological model only after one chooses normalized bases and requires the
unit effect to be represented by the all-ones response functional.  Then
normalized preparations have ontic distributions satisfying
\(\sum_\lambda\mu_\rho(\lambda)=1\) and
\(\xi_u(\lambda)=1\).  In the absence of these order-unit constraints,
Eq.~\eqref{eq:simplex-lp} should be read as a positive cone-factorization
criterion rather than as a full normalized ontological model.

If it is infeasible, one may
introduce an operational noise map
\begin{equation}
        D:\mathcal A\to\mathcal A
\end{equation}
on the state side and minimize $r$ subject to
\begin{equation}
        \begin{aligned}
        I_E^T G\bigl[(1-r)\operatorname{Id}_{\mathcal A}+rD\bigr]I_\Omega
        &= H_E^T\sigma H_\Omega,\\
        \sigma&\ge_{\mathrm e}0,\qquad 0\le r\le1.
        \end{aligned}
        \label{eq:noisy-lp}
\end{equation}
Here $\operatorname{Id}_{\mathcal A}$ denotes the identity map on the ambient
Hermitian-operator coordinate space.  The optimum $r$ is a robustness against
the specified noise.  In the calculations below, $D$ is always the completely
depolarizing state-side map
\begin{equation}
        \rho\longmapsto \tr(\rho)\frac{\id_{d_H}}{d_H},
        \label{eq:depolarizing-map}
\end{equation}
where $d_H=\dim\HH$.  Although Eq.~\eqref{eq:depolarizing-map} is written for positive operators, in Eq.~\eqref{eq:noisy-lp} it is used as its unique linear extension to the ambient real vector space $\mathcal A$ of Hermitian operators.  Accordingly, in the restricted projector-cone audit \(r\) is the robustness of the deformed Born pairing for the labelled fragment.  It should not be identified with an intrinsic robustness of a larger GPT or laboratory scenario unless the noisy preparations and all associated operational equivalences are also included.

Two qualifications fix the scope of the restricted tests used below.
First, the linear program is not a hypergraph-coloring or two-valued-state test.
In the projector-cone version, the context structure is seen only through the
operator geometry of the listed projectors: orthogonality, linear dependence,
and relations such as \(\sum_{v\in C}P_v=\id\) when a complete context is
present.  The program does not impose the deterministic rule that exactly one
atom in each block is true.

Second, simplex embeddability is weaker than simpliciality.  A generalized
probabilistic theory or an accessible fragment may embed into a classical
simplex even though its own state space is not itself a simplex.  This
distinction is central to the generalized-probabilistic-theory analysis of
Schmid \emph{et al.}~\cite{schmid-2021-simplex}.  For accessible fragments,
cone equivalence is especially important: cone-equivalent fragments are either
both classically explainable or both not, because the embedding problem depends
on the generated cones~\cite{selby-2023-fragments}.

Finally, Eq.~\eqref{eq:simplex-lp} is not the same object as the convex hull of
two-valued states of an orthogonality hypergraph.  The two-valued-state hull
appears as a sharp, deterministic shadow of the operational problem when the
fragment contains the eigenstate preparations, sharp projective measurements,
and operational equivalences that force deterministic responses on the relevant
ontic support.  This is the bridge to KS reasoning, not an identity
of the two frameworks.

The computations reported in Table~\ref{tab:results} then impose a deliberately
restricted \emph{projector-cone} version of the simplex test.  The listed
rank-one projectors are used as preparations and as effects, and the cone
generated by those projectors is tested against state-side depolarizing noise;
unit effects, complementary coarse-grainings, and context-induced operational
equivalences are not added unless they are generated by the same projector cone.
Thus two presentations with the same projector set but different groupings into
contexts define the same problem.  The thresholds below are consequently
benchmarks for these projector cones, not for complete operational scenarios.

\section{The orthant-factorization criterion and useful diagnostics}
\label{sec:sandwich}

The exact simplex criterion can be stated compactly as follows.  A finite
prepare-and-measure fragment is classical at the simplex level exactly when its
accessible Born pairing factors through a finite classical orthant.  In the
coordinate convention of Section~\ref{sec:simplex-test}, this Born pairing is
the bilinear form
\begin{equation}
        (y,x)\longmapsto y^T I_E^T G I_\Omega x.
\end{equation}
Thus, after quotienting by operational equivalences, there must exist a finite
ontic set $\Lambda$ and positive linear maps
\begin{equation}
        \mu:S_\Omega\longrightarrow \reals_+^\Lambda,
        \qquad
        \xi:S_E\longrightarrow \reals_+^\Lambda,
\end{equation}
such that, for all included preparations and effects,
\begin{equation}
        \tr(e\rho)
        =
        \sum_{\lambda\in\Lambda}\xi_e(\lambda)\mu_\rho(\lambda),
        \label{eq:orthant-factorization}
\end{equation}
with the normalization and response constraints
\begin{equation}
  \sum_{\lambda}\mu_\rho(\lambda)=1,
  \qquad \xi_u(\lambda)=1,
  \qquad 0\leq\xi_e(\lambda)\leq1
  \label{eq:orthant-normalization}
\end{equation}
for normalized preparations, the unit effect $u$, and included effects $e$.
Thus the classical simplex appears as the normalized slice of the orthant
$\reals_+^\Lambda$.
In this language, simplex nonclassicality is precisely the failure of the
positive factorization in Eq.~\eqref{eq:orthant-factorization} for the specified
fragment.

The orthant may have dimension larger than the accessible operational space.
This possible \emph{dimension gap} is part of the general definition of simplex
embeddability~\cite{schmid-2021-simplex}.  Consequently, the existence of a
simplicial cone $K$ satisfying
$\Cone[\OmegaSet]\subseteq K\subseteq\Cone[\E]^*$ inside the original
accessible state space is a useful sufficient condition, but it is not a
necessary reformulation of Eq.~\eqref{eq:orthant-factorization} unless a
no-dimension-gap assumption is added.  The nonnegative matrix equation
\eqref{eq:simplex-lp}, rather than a same-space simplicial sandwich, is the
general finite criterion used here.  By Farkas duality, infeasibility is
witnessed by a separating linear functional, equivalently by a
noncontextuality inequality for the chosen operational table.

This geometric picture also explains why the simplex test is not a graph invariant.  The orthogonality hypergraph records exclusivity and context incidence, but the simplex linear program sees the accessible cones, the Born bilinear form, and the operational equivalences.  Metric information such as $|\langle\psi|\phi\rangle|^2$ matters.  Conversely, context grouping is invisible to a projector-only cone test unless it is encoded by additional effects, coarse grainings, or equivalence constraints.  The exact question is therefore always whether this specified operational fragment, with these equivalences, admits the positive orthant factorization of Eq.~\eqref{eq:orthant-factorization}.

The following diagnostics are useful, but they should not be mistaken for necessary-and-sufficient graph criteria.  First, a fragment whose accessible state cone is already simplicial is normally classically harmless; nonsimpliciality or overcompleteness of the generated cone is a warning sign.  But merely having many extreme rays is not by itself an obstruction, because the fragment may still factor through a higher-dimensional orthant.  Second, the effects must resolve the nonsimplicial geometry.  Coarse or incomplete effects can make a real obstruction invisible.  Third, the constraints must close a nontrivial consistency loop: cyclic exclusivity, TIFS or KS implication chains, parity structures after spectral refinement, or equal-mixture preparation equivalences.  Without such a loop the hidden-variable sample space often has enough freedom to absorb the transition probabilities.

Three common mechanisms are worth separating.  An \emph{overlap or cyclic-exclusivity obstruction} appears when nonorthogonal quantum states and exclusive effects must be reconciled around a closed cycle; the completed pentagon is the canonical low-dimensional calibration.  A \emph{determinism obstruction} appears when eigenstate preparations and sharp effects force response functions to take values $0$ or $1$ on large parts of the ontic space; in sufficiently interlocked configurations this is the operational shadow of KS, TIFS, and related two-valued-state constraints.  A \emph{convex-decomposition obstruction} appears when different accessible mixtures represent the same operational preparation and hence must be assigned the same ontic distribution.  This last mechanism is the usual preparation-contextuality mechanism and can occur in small scenarios if the relevant mixture equivalences are included.

These mechanisms illuminate the examples below.  The firefly logic \(L_{12}\)~\cite{cohen} realizes complementarity but does not produce a simplex obstruction in the restricted projector-cone fragment~\cite{schaller-92,svozil-2018-b}.  The completed KCBS pentagon already exhibits a cyclic probability-level gap, although it is not KS uncolorable~\cite{Klyachko-2008,wright:pent}.  Tkadlec's nonunital configuration exposes a deterministic two-valued-state defect~\cite{tkadlec-96}.  The original Yu--Oh 13-ray set~\cite{Yu-2012} and its 25-ray orthogonal completion used here, the Cabello--Estebaranz--Garc\'ia-Alcaine 18--9 set~\cite{cabello-96}, and the Peres--Mermin square together with the 24--24 completion~\cite{peres111,mermin90b,peres-91,Pavicic-2005,svozil-2024-convert-pra-externalfigures} give constrained projector-cone fragments whose Born pairings fail simplex embeddability without noise.  GHZ is different: if one retains only the common joint spectral projectors of the four commuting global product observables, the fragment is a single Boolean context and hence simplex-classical.  Its nonclassicality enters through the additional requirement that values of global products factor into pre-existing local Pauli values, a product-preservation constraint not present in the projector-only factorization test~\cite{ghz,mermin90b,svozil-2024-convert-pra-externalfigures}.

\section{Operator-functional tests and masked contexts}
\label{sec:operator-valued}

An ``operator-valued simplex'' is not a single unambiguous object.  Its meaning depends on what is taken as the operational data and what constraints are imposed on the deterministic classical vertices.

First, if an operator is used only as shorthand for its spectral projectors, then there is no essential change from the projector simplex.  A normal observable
\begin{equation}
        A=\sum_i a_iP_i
\end{equation}
contributes expectation values
\begin{equation}
        \langle A\rangle_\rho=\sum_i a_i\,\tr(\rho P_i),
\end{equation}
which are linear post-processings of the spectral-outcome probabilities.  If the projectors $P_i$ are already included as effects, adding $A$ itself changes only the coordinates of the data table.

Second, if one keeps only a degenerate operator and does not include enough commuting observables to resolve its eigenspaces, then the simplex test becomes coarser.  Some distinctions between projectors are hidden.  This can make simplex embeddability easier and may miss contextuality that appears after maximal refinement.  The matrix-pencil method addresses precisely this issue: a suitable linear combination of mutually commuting degenerate operators reveals the common spectral projectors and hence the underlying context~\cite{svozil-2024-convert-pra-externalfigures}.

Third, if the classical model is required to assign values to operators while preserving functional and product relations, the test changes substantially.  The deterministic vertices are no longer arbitrary assignments to spectral outcomes; they must also satisfy relations such as Eq.~\eqref{eq:product-rule}.  This is the setting of Peres--Mermin and GHZ parity arguments.  In such cases the obstruction may be algebraic even when a projector-only simplex for a single context would be trivial.

Consequently, an operator-valued simplex can mean at least three different things: a simplex for expectation values, a simplex for spectral-outcome probabilities, or a simplex whose vertices are deterministic operator value assignments constrained by functional calculus and, where applicable, by multiplicative product rules.  The first two are operational probability tests.  The last is an operator-functional test, and the GHZ ingredient is specifically the product-rule part, not merely functional calculus on a single spectral resolution.  This is the sense in which GHZ, although a single context after diagonalization, still violates classical expectations: the violation is not in the Boolean algebra of its joint spectral projectors but in the attempted decomposition of its global product observables into context-independent local elements of reality.

\subsection{Consecutive global measurements versus local factorizations}
\label{subsec:ghz-consecutive}

One might object that the four GHZ product observables commute.  Why not simply measure them consecutively on the same three-particle system?  In notation, the four observables are
\begin{align}
        A_1&=X_1X_2X_3,&
        A_2&=X_1Y_2Y_3,\nonumber\\
        A_3&=Y_1X_2Y_3,&
        A_4&=Y_1Y_2X_3 .
\end{align}
This consecutive protocol is possible in principle.  Since the $A_i$ commute, their spectral projections commute.  An ideal L\"uders measurement of one product observable collapses the state only to an eigenspace that is invariant under all the others.  Subsequent ideal measurements refine the state to a common eigenspace and preserve the registered eigenvalues.  Equivalently, one may perform the joint projective measurement with atoms
\begin{equation}
        P_{\boldsymbol\varepsilon}
        =\prod_{i=1}^4 \frac{\id+\varepsilon_i A_i}{2},
        \qquad
        \varepsilon_i\in\{\pm1\},
        \qquad
        \varepsilon_1\varepsilon_2\varepsilon_3\varepsilon_4=-1,
        \label{eq:ghz-joint-pvm}
\end{equation}
where only eight sign patterns are nonzero because $A_1A_2A_3A_4=-\id$.
Only three signs are independent; the fourth is fixed by the product
constraint.  Thus a collective consecutive protocol registers four repeatable
outcomes whose product is always $-1$.

\begin{remark}[Commuting observables and registered outcomes]\label{rem:commuting-lueders}
The preceding statement uses the standard finite-dimensional spectral fact that
commuting self-adjoint observables admit compatible L\"uders measurements.
Let $A$ and $B$ be self-adjoint operators with $[A,B]=0$, and let
$P_a$ and $Q_b$ be their spectral projectors.  Since spectral projectors
are obtained from the operators by functional calculus, $[P_a,Q_b]=0$ for
all eigenvalues $a,b$.  Hence $B P_a=P_aB$, so the subspace
$P_a\HH$ is invariant under $B$.  Equivalently,
\begin{equation}
        P_a\HH=\bigoplus_b P_aQ_b\HH,
        \label{eq:joint-eigenspace-refinement}
\end{equation}
up to zero summands.  Thus the collapse caused by registering the outcome
$a$ of $A$ does not cut across the eigenspace decomposition of $B$; it only
selects a direct sum of joint eigenspaces.

For a L\"uders measurement of $A$ followed by a measurement of $B$, the
sequential joint probability is
\[
        p(a\ \mathrm{then}\ b)
         =\tr\bigl(Q_b P_a\rho P_a\bigr)
         =\tr\bigl(P_aQ_b\rho\bigr).
\]
which is precisely the probability assigned by the simultaneous joint PVM
$\{P_aQ_b\}_{a,b}$.  A later measurement of $B$ may refine the post-measurement
state within $P_a\HH$, but it cannot invalidate the registered
value $a$.  This proof applies to the degenerate observable itself.  It
does not apply to an arbitrary finer apparatus that, while reporting the
same coarse outcome $a$, measures additional degrees of freedom inside
$P_a\HH$ which need not commute with $B$.
\end{remark}

This collective protocol does not, by itself, yield a GHZ contradiction.  It is just one joint measurement, hence one Boolean context.  A classical hidden-variable model for it is immediate: the ontic state is one of the eight allowed sign patterns $\boldsymbol\varepsilon=(\varepsilon_1,\varepsilon_2,\varepsilon_3,\varepsilon_4)$ satisfying $\prod_i\varepsilon_i=-1$, and the measurement reads off its four coordinates.  No product rule for local factors is used, and no inconsistency follows.

The GHZ contradiction appears only if the same global product values are required to arise from pre-existing local Pauli values,
$v(X_i),v(Y_i)\in\{\pm1\}$,
$v(X_1Y_2Y_3)=v(X_1)v(Y_2)v(Y_3)$,
with analogous equations for the other three products.  Multiplying the four classical equations gives $+1$, because every local factor occurs twice.  Quantum mechanically the product of the four commuting global operators is $-\id$.  The collective measurement of Eq.~\eqref{eq:ghz-joint-pvm} reveals the global products, but it never reveals the six local quantities $X_i,Y_i$ whose context-independent coexistence is being assumed.  In that precise sense the collective protocol is factor-blind.

The local GHZ protocol operationalizes a different refinement.  In each run the three separated parties choose one of the four compatible triples
        $\{X_1,X_2,X_3\}$,
        $\{X_1,Y_2,Y_3\}$,
        $\{Y_1,X_2,Y_3\}$,
        $\{Y_1,Y_2,X_3\}$,
and multiply the local outcomes.  These four triples are not jointly measurable by local sharp measurements, since on each particle $X_i$ and $Y_i$ anticommute.  Locality, or equivalently the Einstein--Podolsky--Rosen (EPR) counterfactual step in the usual GHZ reasoning, is what motivates assigning the same value to $X_i$ or $Y_i$ independently of which compatible triple is chosen.  The contradiction therefore concerns the attempted identification of two operationally different realizations of a product observable: a collective entangled measurement of the product as a whole, and a local measurement of its factors followed by multiplication.

There is a final technical caveat.  A degenerate observable does not determine a unique measurement apparatus.  A badly chosen implementation may disturb states inside a degenerate eigenspace by effectively measuring additional, possibly incompatible, degrees of freedom.  That is not a failure of commutativity of the abstract observables; it is a different physical refinement of the degenerate measurement.  For the ideal consecutive protocol considered here, one assumes L\"uders instruments or the common joint PVM, in which case the registered product outcomes are compatible and repeatable.  What remains nonclassical is not the sequence of global product outcomes, but the extra product-preserving identification of those outcomes with jointly pre-existing local values.

\subsection{Single-context triviality, unmasking, and the chromatic analogy}
\label{subsec:ghz-unmasking}

The preceding discussion also fixes a possible misconception about the scope of simplex tests.  A closed fragment consisting of one joint spectral measurement is always simplex embeddable.  Its ontic states may simply be the atoms of that Boolean algebra, and an arbitrary preparation is represented by the corresponding probability distribution over those atoms.  Thus the collective GHZ product fragment, taken by itself, has no projector-simplex obstruction.  This is not a positive classical explanation of the full GHZ experiment; it is only the trivial simplex embeddability of a closed single Boolean context.

The qualifier is essential.  One should not conclude that operator-valued arguments can never enter a simplex analysis.  They can do so whenever the operator proof is unfolded into a multi-context projector fragment, or whenever the operational scenario includes the local factors and their compatibility relations.  Peres--Mermin supplies the clearest example: its algebraic parity proof, once the degenerate observables are resolved by simultaneous diagonalization or matrix pencils, yields a genuine multi-context projective configuration, and that configuration can be subjected to two-valued-state and simplex tests.  The GHZ collective context does not itself have this property; the local GHZ experiment has a different operational structure, namely four incompatible local settings together with the product-rule identification of shared local factors.

The operative distinction is therefore not whether an argument is operator-valued or simplex-theoretic in the abstract.  The right question is whether the retained fragment is merely a single collective spectral context or an unmasked multi-context fragment.  A single collective context is classically representable as a Boolean block.  A multi-context fragment may display ordinary projector contextuality, simplex nonembeddability, Bell--GHZ nonlocality, or an operator-functional parity contradiction, depending on which structures are retained in the fragment.

This also clarifies the relation to chromatic contextuality.  GHZ and strong chromatic contextuality are similar only in the negative sense that neither is exhausted by a probability-table simplex test.  Positively they are different obstructions.  Chromatic contextuality is combinatorial: it asks whether a fixed nondegenerate spectrum, or equivalently a strong coloring, can be assigned globally across a hypergraph.  GHZ is algebraic: it asks whether values can preserve products of commuting observables and their local factors.  State-independent parity proofs such as Peres--Mermin are closer in spirit to chromatic obstructions because they express a global spectral inconsistency independent of the input state, but the mechanism is still operator multiplication rather than hypergraph coloring.  Thus both belong outside the simplex-probability axis, but they lie on different non-simplex axes: one incidence/coloring-theoretic, the other operator-algebraic.

\section{Why two contexts need not, by themselves, reveal nonclassicality}
\label{sec:two-contexts}

Before turning to calibration examples, it is useful to remove one smaller temptation: the idea that the passage from one context to two already separates quantum vector probabilities from classical convex probabilities.  That intuition is correct at the level of the full theories but misleading for a small operational fragment.  A single orthonormal context is an ordinary probability simplex.  Two nonintertwining contexts are, operationally, a pair of simplexes connected by a stochastic transition matrix.

Let
\begin{equation}
        A=\{a_1,\ldots,a_d\},\qquad B=\{b_1,\ldots,b_d\}
\end{equation}
be two orthonormal bases.  The quantum transition matrix
\begin{equation}
        T_{ji}=|\langle b_j|a_i\rangle|^2
\end{equation}
is doubly stochastic.  For the restricted fragment consisting only of preparations in $A\cup B$ and measurements in $A\cup B$, this matrix can be reproduced by a classical random channel.  One may take ontic states to be pairs $\lambda=(i,j)$, where $i$ is the predetermined answer to context $A$ and $j$ is the predetermined answer to context $B$.  Preparing $a_i$ means loading the classical ensemble with the states $(i,1),\ldots,(i,d)$ in proportions $T_{1i},\ldots,T_{di}$.  A measurement of $A$ reads the first coordinate and therefore returns $i$ with certainty; a measurement of $B$ reads the second coordinate and returns the Born transition probabilities.  Preparations $b_j$ are treated symmetrically.  No independence assumption is being imposed on two physical subsystems here; $\lambda=(i,j)$ is only a bookkeeping device for predetermined answers in two otherwise disconnected measurement contexts.  The construction works precisely because no further cross-context consistency constraints are present.

If two contexts share an atom, the same pair-coordinate construction must be
quotiented so that the shared effect has the same response in both contexts.
For weak pastings such as \(L_{12}\), this quotienting is exactly what the
partition-logic representation accomplishes.

This construction is not a classical simulation of the full Hilbert-space
theory, nor of an arbitrary multi-context or Bell scenario.  In a Bell
experiment a single hidden-variable distribution and a single family of local
response functions must work for every local setting; setting-by-setting fits
that cannot be assembled into such a common factorization are not Bell-local
models.  Likewise, additional preparation equivalences, measurement
equivalences, linear dependences among preparations or effects, or a third
context can already obstruct a common simplex embedding.  The limited claim
here is only that a bare transition table between two bases, with no further
operational constraints beyond normalization, has a classical stochastic-channel
realization.  Nonclassicality becomes visible when one asks for a single model
to satisfy enough cross-context constraints around a nontrivial web of pasted
contexts.

This is precisely the lesson of partition logics.  A partition logic is obtained by pasting Boolean algebras arising from partitions of a finite set.  Its atoms can be represented as subsets of a common classical sample space, and probabilities are ordinary convex mixtures over dispersion-free states.  Equivalently, the same structures can be modeled by Wright-style generalized urns or by the initial-state identification problem for finite deterministic Moore or Mealy automata.  In a generalized urn, a ball type is an ontic state and different colors correspond to incompatible experimental contexts; looking through one color filter reveals one partition while hiding the others.  In the automaton model, the initial state is the ontic state and different input symbols induce different observable partitions of the state set.  These models realize complementarity without value indefiniteness.

The contrast with a Hilbert-space realization is then not in the exclusivity graph alone.  The same graph may support a set-theoretic partition representation, a faithful orthogonal vector representation, or even more exotic weights~\cite{svozil-2018-b}.  The probability type is determined by the chosen physical representation and the operational fragment, not by the bare pasting diagram alone.  Thus the statement ``complementarity does not imply contextuality'' is not merely philosophical: it is a concrete modeling fact.

\section{Calibration examples: firefly logic and the pentagon}
\label{sec:calibration}

The first calibration example is the smallest nontrivial pasting.  The firefly logic $L_{12}$ is a 5--2 vertex--context set, obtained by pasting two three-atomic Boolean algebras along one common atom~\cite{cohen},
\begin{equation}
        \{a,b,c\},\qquad \{a,d,e\}.
\end{equation}
A faithful qutrit realization is
\begin{equation}
        a=(1,0,0),\quad b=(0,1,0),\quad c=(0,0,1),
\end{equation}
\begin{equation}
        d=(0,1,1),\quad e=(0,1,-1).
\end{equation}
The partition-logic representation is even more elementary.  On the finite set $S_5=\{1,\ldots,5\}$ one may use the two partitions
\begin{equation}
        \bigl\{\{2,3\},\{4,5\},\{1\}\bigr\},\qquad
        \bigl\{\{1\},\{3,5\},\{2,4\}\bigr\},
        \label{eq:l12-partition}
\end{equation}
which paste along the singleton atom $\{1\}$.  This is a generalized urn or finite-automaton realization of the same logic.  It is non-Boolean as a pasted structure, but it is still classically representable.  Its role here is diagnostic: it is the first place where contexts are not simply disjoint, yet simplex nonembeddability should not be expected merely from that fact.

The next calibration example is the pentagon, or pentagram logic.  In the KCBS form one starts with five qutrit rays $v_i$ satisfying cyclic orthogonality
\begin{equation}
        v_i\perp v_{i+1}\qquad (i\;\mathrm{mod}\;5).
\end{equation}
A convenient set of unnormalized ray representatives is
\begin{equation}
        v_k\propto\left(
        \cos\frac{4\pi k}{5},\;
        \sin\frac{4\pi k}{5},\;
        \sqrt{\cos\frac{\pi}{5}}
        \right),\qquad k=0,\ldots,4.
        \label{eq:kcbs-rays}
\end{equation}
Completing each orthogonal pair by $w_i=v_i\times v_{i+1}$ gives five complete qutrit contexts
\begin{equation}
        \{v_i,w_i,v_{i+1}\},\qquad i=0,\ldots,4,
        \label{eq:pentagon-contexts}
\end{equation}
and hence a 10-ray pentagon (10--5 vertex--context set).

The pentagon is still partition-logically representable: the five cyclically intertwined contexts support 11 two-valued states, from which a partition logic over $S_{11}=\{1,\ldots,11\}$ can be reconstructed; explicitly, one may take the 11 admissible two-valued states as the underlying points and represent each atom by the subset of states assigning value one to it.  Nevertheless, the pentagon already displays a difference between classical convex probabilities and Born/Lov\'asz-type vector probabilities.  On the five outer atoms, the classical convex hull yields the familiar pentagon bound $\sum_i p_i\le2$, whereas maximization over quantum states for the symmetric orthogonal representation gives $\sqrt5$.  The numerical threshold reported below belongs to the symmetric algebraic realization in Eq.~\eqref{eq:kcbs-rays}; it is not a graph invariant of \(C_5\) alone.  This is the first useful calibration point at which cyclic exclusivity exposes a quantitative difference, though still not a KS impossibility.

\section{Combinatorial contextuality: two-valued states and meagerness}

Let $H=(V,\C)$ be a finite $d$-uniform orthogonality hypergraph.  The vertices $v\in V$ represent atoms, usually rank-one projectors, and each context $C\in\C$ is a $d$-element set of mutually orthogonal atoms corresponding to a maximal sharp measurement.  A faithful orthogonal representation in $\complex^d$ or $\reals^d$ assigns to every vertex a ray $|v\rangle$ such that vertices are orthogonal precisely when the hypergraph says they are, and every context resolves the identity:
\begin{equation}
        \sum_{v\in C} \Pi_v = \id,
        \qquad \Pi_v=\frac{|v\rangle\langle v|}{\langle v|v\rangle}.
\end{equation}

\begin{definition}[Two-valued state]
A two-valued state on $H$ is a map $s:V\to\{0,1\}$ such that
\begin{equation}
        \sum_{v\in C}s(v)=1
        \qquad \text{for every } C\in\C.
        \label{eq:tvs}
\end{equation}
We denote the set of all such states by $\TVS(H)$.
\end{definition}

The deterministic valuation polytope is
\begin{equation}
        \Ptwo(H)=\conv\TVS(H)\subseteq[0,1]^V.
\end{equation}
It is useful to distinguish several ways in which $\TVS(H)$ can be deficient.  Tkadlec called attention to hypergraphs whose state spaces are empty, not unital, not separating, or not full~\cite{tkadlec-96}; these notions are standard in the quantum-logic literature~\cite{pulmannova-91,schaller-92,svozil-tkadlec}.

\begin{definition}[Unital, separating, full]
Let $\TVS(H)$ be the set of two-valued states on $H$.
It is called
\begin{enumerate}
\item \emph{unital} if for every vertex $a\in V$ there exists $s\in\TVS(H)$ with $s(a)=1$;
\item \emph{separating} if for every distinct pair $a,b\in V$ there exists $s\in\TVS(H)$ with $s(a)\ne s(b)$;
\item \emph{full} if for every nonorthogonal pair $a,b\in V$ there exists $s\in\TVS(H)$ with $s(a)=s(b)=1$.
\end{enumerate}
In the usual orthomodular-poset formulation, where atoms sit inside a poset with zero, unit, and orthocomplementation, fullness implies separation and separation implies unitality.  In the atom-only hypergraph formulation used here these implications require the corresponding reconstruction assumptions.  We therefore treat unitality, separation, and fullness as separate diagnostics and do not rely on the hierarchy without those assumptions.
\end{definition}

\begin{definition}[Meager two-valued-state space]
A hypergraph has a meager two-valued-state space if $\TVS(H)$ is empty, nonunital, nonseparating, or nonfull.  This is an umbrella term; the four cases have different operational meanings.
\end{definition}

Each failure mode corresponds to a different linear constraint on $\Ptwo(H)$.  If $\TVS(H)=\emptyset$, then $\Ptwo(H)=\emptyset$.  This is the KS case.  If $\TVS(H)$ is nonunital at $a$, then
\begin{equation}
        x_a=0 \qquad \text{for all }x\in\Ptwo(H).
        \label{eq:nonunital-face}
\end{equation}
If it is nonseparating for $a\ne b$, meaning $s(a)=s(b)$ for all $s\in\TVS(H)$, then
\begin{equation}
        x_a=x_b \qquad \text{for all }x\in\Ptwo(H).
        \label{eq:nonsep-face}
\end{equation}
If it is nonfull for a nonorthogonal pair $a,b$, meaning no $s$ has $s(a)=s(b)=1$, then
\begin{equation}
        x_a+x_b\le1 \qquad \text{for all }x\in\Ptwo(H).
        \label{eq:nonfull-face}
\end{equation}
A true-implies-false (TIFS) structure, such as the Specker bug, singles out an ordered input--output pair $a,b$ for which
\begin{equation}
        s(a)=1\Longrightarrow s(b)=0,
        \label{eq:tifs}
\end{equation}
for all admissible $s$.  At the level of binary assignments this is equivalent to the nonfullness inequality $s(a)+s(b)\leq1$; the direction enters through the distinguished preparation or conditioning on $a$, not through a stronger binary constraint.  Equation~\eqref{eq:tifs} implies that the face $x_a=1$ of the two-valued-state polytope is contained in $x_b=0$.

\section{How meagerness becomes operational}

The preceding notions are combinatorial: they concern the possible two-valued
states on a hypergraph.  To become operational, however, such defects must be
visible in the actual preparations and effects included in the fragment.  A
logical obstruction that is present in the hypergraph may remain invisible if
the fragment is too poor to test it.

Let $H$ have a faithful orthogonal representation, with vertex projectors
$\Pi_v$.  For a preparation $\rho$, write
\begin{equation}
        p^\rho_v=\tr(\rho\Pi_v)
\end{equation}
for the Born probability assigned to the vertex effect $\Pi_v$.  The
deterministic two-valued-state polytope
\begin{equation}
        \Ptwo(H)=\conv\TVS(H)
\end{equation}
contains exactly the probability assignments obtainable as classical mixtures
of admissible two-valued states.  Thus, if an accessible Born vector
$p^\rho$ lies outside $\Ptwo(H)$, then no deterministic two-valued-state
model can reproduce that part of the fragment.

The different forms of meagerness become visible in different ways.
If the two-valued states are nonunital at an atom $a$, then every admissible
two-valued state assigns
\begin{equation}
        s(a)=0.
\end{equation}
Consequently every classical mixture in $\Ptwo(H)$ also assigns probability
zero to $a$.  But if the fragment allows the eigenstate preparation
$\rho=\Pi_a$ and the effect $\Pi_a$, quantum theory gives
\begin{equation}
        \tr(\Pi_a\Pi_a)=1.
\end{equation}
The atom that is never true in the two-valued-state model is obtained with
certainty in the corresponding quantum eigenstate.  Nonunitality is then
directly operationally visible.

If the two-valued states are nonseparating for two distinct atoms $a$ and
$b$, then every admissible valuation identifies them:
\begin{equation}
        s(a)=s(b).
\end{equation}
Every classical mixture therefore satisfies $x_a=x_b$.  This becomes
operationally relevant only if the Born statistics can distinguish the two
effects, that is, if the fragment contains some preparation $\rho$ such that
\begin{equation}
        \tr(\rho\Pi_a)\ne \tr(\rho\Pi_b).
\end{equation}
In that case the classical valuation model merges two outcomes that the quantum
statistics separate.  If no such preparation is included, the nonseparability
remains present combinatorially but is not witnessed by the fragment.

Finally, a true-implies-false (TIFS) or nonfullness constraint becomes visible
through nonorthogonal overlap.  Suppose $a$ and $b$ are nonorthogonal, but
the two-valued states force
\begin{equation}
        s(a)=1\Longrightarrow s(b)=0.
\end{equation}
Classically, conditioning on $a$ being true then forces $b$ to be false.
Quantum mechanically, however, the eigenstate $\Pi_a$ gives
\begin{equation}
        \tr(\Pi_a\Pi_b)=|\langle a|b\rangle|^2>0
\end{equation}
whenever $a$ and $b$ are nonorthogonal.  Thus, if the fragment contains the
preparation $\Pi_a$ and the effects $\Pi_a,\Pi_b$, the TIFS implication is
operationally contradicted by the nonzero transition probability from $a$ to
$b$.

The qualification is essential.  These are direct witnesses against models
whose ontic states are the admissible deterministic two-valued states.  They
become witnesses against generalized-noncontextual models only when the
operational scenario justifies the corresponding outcome-determinism and
equivalence assumptions for the sharp effects.  If a nonunital atom is never
prepared or measured, if a nonseparated pair is never statistically
distinguished, or if the endpoints of a TIFS gadget are absent from the
preparation/effect list, then the combinatorial defect need not appear in the
operational table.  The noisy cone-factorization LP then quantifies the
robustness of the stated fragment, rather than automatically that of every
operational realization of the hypergraph.

\section{Relation to generalized noncontextuality}

The deterministic two-valued-state analysis is closest to the traditional KS
setting.  Generalized noncontextuality is broader and does not assume
determinism from the outset.  For sharp projective measurements, the relevant
deterministic constraints can be derived only under the appropriate ideal
operational assumptions, including the projective measurements and the
preparation or mixture equivalences used in the derivation
\cite{kunjwal-spekkens-2015}.  The valuation polytopes below should therefore
be read as deterministic shadows of suitably specified operational models, not
as a substitute for stating those assumptions.

For a fixed accessible GPT fragment with all operational equivalences already
quotient\-ed out, the factorization Eq.~\eqref{eq:simplex-lp} is the
corresponding cone form of the simplex embedding problem.  The computations in
this paper choose a smaller fragment, so the resulting \(r\)-values are
robustnesses of that fragment, not automatic lower or upper bounds on a
separately specified full operational robustness.  The closure computation in
Table~\ref{tab:closure-results} is one intermediate audit: it restores the unit,
residual sharp-measurement outcomes, complements, coarse grainings, and some
preparation equivalences generated by the same vector data.  It is still not a
claim to enumerate every possible laboratory realization of the scenario.

In this situation the two-valued-state polytope $\Ptwo(H)$ is the deterministic shadow of the operational simplex model.  Meagerness of $\TVS(H)$ supplies faces or equalities that the Born vectors may violate.  Depolarizing noise moves the Born vectors toward a more symmetric point, and the noisy simplex linear program computes the amount of noise needed to make the entire fragment simplex-embeddable.

The crucial qualification is that the simplex linear program is not merely ``the fractional coloring polytope.''  Its variables are the entries of $\sigma$ in Eq.~\eqref{eq:simplex-lp}, indexed by facets of the state and effect cones.  The number of such facets need not equal the number of vertices or contexts of the hypergraph.  In the Yu--Oh 13-ray computation below, there are 13 projectors but 24 facets of each cone, hence $\sigma$ is $24\times 24$.

\section{Combinatorial contextuality: chromatic completeness}

Chromatic contextuality imposes a different global requirement.  Let $H=(V,\C)$ be $d$-uniform.  A strong $d$-coloring is a map
\begin{equation}
        c:V\to\{1,\ldots,d\}
\end{equation}
such that every context contains every color exactly once.  Equivalently, for every color $k$ define
\begin{equation}
        s_k(v)=\begin{cases}
        1, & c(v)=k,\\
        0, & c(v)\ne k.
        \end{cases}
\end{equation}
Then each $s_k$ is a two-valued state and
\begin{equation}
        \sum_{k=1}^d s_k(v)=1
        \qquad \text{for every }v\in V.
        \label{eq:color-decomp}
\end{equation}
Thus chromatic colorability asks not merely for two-valued states, but for a complete decomposition of the unit assignment into $d$ admissible two-valued states.  The obstruction $\chi_s(H)>d$, where $\chi_s$ is the strong chromatic number, means that no such global spectral-label decomposition exists.

This explains why chromatic contextuality is independent of ordinary valuation scarcity.  A hypergraph may possess a separating or even full set of two-valued states and nevertheless fail Eq.~\eqref{eq:color-decomp}.  Conversely, a KS hypergraph with no two-valued states is of course chromatically contextual, but the chromatic description is then not the most discriminating one: the stronger valuation obstruction is already present.

\section{Computational protocol}

The computational analysis fixes one convention for all examples.  The listed
rays define rank-one projectors, these projectors are used both as preparations
and as effects, and depolarizing noise is applied only on the state side.  The
named hypergraphs determine the projector sets and are used in the accompanying
combinatorial analysis; in the linear program, their blocks enter only through the operator
geometry of the included projectors.

The computations use the non-orthonormal coordinate convention described in
Section~\ref{sec:simplex-test}.  For exact rational audits the program represents
real symmetric projectors in the raw coordinate basis
\begin{equation}
        (\rho_{11},\ldots,\rho_{dd},
        \rho_{12},\rho_{13},\ldots,\rho_{d-1,d}),
        \label{eq:raw-basis}
\end{equation}
rather than in a trace-orthonormal basis.  In this basis the trace pairing is
represented by the diagonal matrix
\begin{equation}
        G=\operatorname{diag}
        \Big(
        \underbrace{1,\ldots,1}_{d},
        \underbrace{2,\ldots,2}_{d(d-1)/2}
        \Big),
        \label{eq:raw-trace-metric}
\end{equation}
because the \(d\) diagonal entries contribute once whereas the \(d(d-1)/2\)
off-diagonal real symmetric entries contribute twice to the trace
inner product.  Thus, if $e$ and $\rho$ denote raw coordinate vectors, then
\begin{equation}
        \tr(e\rho)=e^T G\rho.
        \label{eq:raw-trace-pairing}
\end{equation}
The consolidated audit stores the dual effect vector $\widetilde e=Ge$, whose
ordinary dot product with $\rho$ gives the same Born probability.  This convention avoids
square-root factors for integer rays and makes the rational cases suitable for
exact polyhedral conversion with \texttt{pycddlib}/cddlib over $\mathbb Q$.

The audit uses the accessible-fragment formalism.  Starting from full raw
state coordinates $V^F_\Omega$ and full dual-effect coordinates
$\widetilde V^F_E$, Stage~1 replaces them by accessible coordinate matrices
$V^A_\Omega$ and $\widetilde V^A_E$ of full row rank, together with embedding
matrices $I_\Omega$ and $\widetilde I_E$.  When the operational
quotient is nontrivial, these matrices represent the chosen linear sections;
otherwise they are the ordinary inclusions.  In either case they embed the
chosen accessible representatives back into the full raw coordinate space, so
that
\begin{equation}
        V^F_\Omega=I_\Omega V^A_\Omega,
        \qquad
        \widetilde V^F_E=\widetilde I_E\widetilde V^A_E
\end{equation}
up to the chosen basis convention.
Stage~2 computes facet descriptions
$H_\Omega$ and $H_E$ of the accessible cones.
Because the metric has already been absorbed into the dual-effect coordinates,
Stage~3 solves
\begin{equation}
        \widetilde I_E^T\bigl[(1-r)\operatorname{Id}_{\mathcal A}+rD\bigr]I_\Omega
        =
        H_E^T\sigma H_\Omega,
        \,
        \sigma\ge_{\mathrm e}0,
        \,
        0\le r\le1.
        \label{eq:stage3-audit}
\end{equation}
This is Eq.~\eqref{eq:noisy-lp} after the accessible reduction and the
substitution $\widetilde I_E=GI_E$.  Stating this substitution explicitly
prevents the trace metric from being counted twice.

Candidate optima are found numerically.  For rows reported as exact optima,
the proposed values are then checked symbolically with rational
$H_\Omega,H_E,r$, and $\sigma$, and exact rational dual witnesses are verified.

The certificate status in Table~\ref{tab:results} distinguishes two cases.
``Exact optimum'' means that both an exact primal certificate and an exact dual
certificate have been verified.  ``Numerical'' means that only a floating-point
solution and its residual or dual-feasibility check are available.  The
completed pentagon is algebraic rather than rational in the chosen coordinates
and is therefore treated by a numerical branch.  The Specker-bug coordinates
belong to $\mathbb Q(\sqrt2)$, but their Born table admits an exact rational
rank factorization; exact primal and dual witnesses certify $r=25/79$.
The Kochen--Specker $\Gamma_3$ Born table is genuinely quadratic and is treated
numerically.  The Tkadlec nonunital row has rational coordinates and exact
facets, but no exact primal--dual pair was recovered at its numerical optimum.

For the numerical rows, the printed digits are descriptive rather than
certified and should be read together with the residual information in
Appendix~\ref{app:audit}.

The symmetric choice used in this audit,
\[
        \Omega=\mathcal E=\{\Pi_v:v\in R\},
\]
is not part of the simplex-embeddability criterion itself.  The general
criterion allows the preparation and effect cones to be different.  The present
choice is a benchmarking convention: every listed ray is assumed to be both
sharply preparable and sharply measurable, so that the Born table contains all
projector overlaps \(\operatorname{tr}(\Pi_v\Pi_w)\).  If instead one used only
a single orthonormal basis of preparations, the resulting fragment would usually
be simplex-embeddable.  Indeed, for preparations
\(\rho_k=|k\rangle\langle k|\) and arbitrary projective effects
\(\Pi_v=|v\rangle\langle v|\), the classical model with ontic states
\(\lambda=k\), preparation distributions
\(\mu_k(\lambda)=\delta_{\lambda k}\), and response functions
\(\xi_v(\lambda)=|\langle \lambda|v\rangle|^2\) reproduces
\[
        \operatorname{tr}(\Pi_v\rho_k)
        =
        \sum_\lambda \xi_v(\lambda)\mu_k(\lambda).
\]
Thus such a preparation-restricted fragment cannot by itself witness the
valuation scarcity of the underlying ray hypergraph.

The second audit uses the same vector coordinates but replaces the
projector-only effect set by a vector-generated operational closure.  For every
mutually orthogonal subset of the listed rays, the program forms the sharp
measurement consisting of those rank-one projectors together with the residual
effect \(\id-\sum_i\Pi_i\), whenever this residual is nonzero.  It then adds
the unit effect and all nonempty proper coarse grainings, and identifies equal
operators in the raw coordinate quotient.  Two preparation variants are
reported.  The \emph{closed-effects} variant keeps the original eigenstate
preparations and adds the maximally mixed state.  The \emph{closed-effects plus
preparations} variant also adds normalized versions of the closed measurement
events as preparations whenever their trace is nonzero.  These closure rows are
computed from exact rational facets when possible, but the values in
Table~\ref{tab:closure-results} are numerical unless an exact primal and dual
certificate is explicitly stated.  For the large $\Gamma_3$ and Tkadlec
nonunital closure rows, the
linear program was solved in a cutting-dual form: this gives the same numerical
optimum but does not produce a primal \(\sigma\) certificate.

All calculations reported here are produced by one self-contained program,
\texttt{simplex\_audit\_rebuilt.py}.  Before constructing any cone or LP, it
checks projective uniqueness, enumerates all orthogonal contexts, and verifies
the advertised valuation properties.  The source-derived Specker bug has 13
rays, seven triads, 14 two-valued states, and the distinguished nonfull pair.
The $\Gamma_3$ construction of Kochen and Specker~\cite{kochen1}, in the
coordinate-labelled realization of Tkadlec~\cite{tkadlec-96}, is completed to
27 rays and 17 triads; its 24 two-valued states identify the distinguished
nonseparating pair.  The Tkadlec nonunital construction starts from the 25
marked rays, completes every orthogonal pair by its unique cross-product ray,
and verifies 37 rays, 26 triads, eight two-valued states, and eight never-true
rays.  Cabello's thesis gives an independent parametric coordinatization of
the Bell/Specker TIFS family~\cite{Cabello-1996-diss}.  The Peres--Mermin rays are checked against the six
commuting operator contexts of the square.  The machine-readable report stores
the vectors, contexts, cone generators, facets, LP matrices, numerical
solutions, and all certificates.

OpenAI Codex with a GPT-5.6-based model (July 2026) assisted in consolidating and
reviewing the audit code and in revising the exposition.  The author specified
the mathematical constructions, inspected the generated code, executed all
reported runs, and checked the outputs.  Exact claims are accepted by the
program only after exact rational primal and dual verification; otherwise they
remain explicitly numerical.

\section{Examples and results}

\subsection{Calibration computations: \texorpdfstring{$L_{12}$}{L12} and the completed pentagon}

The firefly logic $L_{12}$ is included as a calibration for complementarity without simplex obstruction.  Its five projectors span a four-dimensional accessible state space inside the six-dimensional real symmetric qutrit trace space.  After the Stage~1 accessible-fragment reduction the exact rational audit finds five facets for each accessible cone and returns
\begin{equation}
        r_{L_{12}}=0,
\end{equation}
with both exact primal and exact dual certificates.  Thus $L_{12}$ is not merely a lower-rank exception to the full-rank trace-space code; it is an exactly simplex-embeddable accessible fragment.  This result is consistent with the partition representation of its incidence structure, but it does not follow from that representation alone: a partition-logically representable pasting can support a different quantum Born table, as the pentagon comparison illustrates.

The completed pentagon contains 10 projectors and spans the full six-dimensional real symmetric qutrit trace space.  The computation verifies adjacent orthogonality, obtains five contexts of the form Eq.~\eqref{eq:pentagon-contexts}, finds 12 cone facets, and returns
\begin{equation}
        r_{C_5}\simeq0.119140568134.
        \label{eq:pentagon-result}
\end{equation}
The value is numerical only in the present audit because the coordinates
involve algebraic quantities outside the rational cddlib backend.  It is a
positive value for this particular projector-cone proxy; it should not be read
as a strength ranking against the later configurations.

\subsection{Specker bug and the Kochen--Specker \texorpdfstring{$\Gamma_3$}{Gamma3} configuration}

The Specker bug is represented by 13 rays in seven triads.  The coordinate
realization uses entries in $\mathbb Q(\sqrt2)$ and is independently supported
by the Bell/Specker parametrization in Cabello's thesis~\cite{Cabello-1996-diss}
and by the coordinate-labelled diagram of Tkadlec~\cite{tkadlec-96}.  Exact
enumeration gives 14 two-valued states and verifies the distinguished nonfull
pair, equivalently the TIFS implication after conditioning on its input ray.
Although the ray coordinates are quadratic, every Born overlap belongs to
$\mathbb Q$.  The program therefore constructs an exact rational rank
factorization of the Born table.  The state and effect cones each have 26
facets, and exact primal and dual certificates give
\begin{equation}
        r_{\mathrm{bug}}=\frac{25}{79}\simeq0.316455696203.
        \label{eq:specker-bug-result}
\end{equation}

The configuration denoted $\Gamma_3$ in the Kochen--Specker construction~\cite{kochen1}
is evaluated using the coordinate-labelled realization recorded by
Tkadlec~\cite{tkadlec-96}.  Orthogonal completion gives 27 rays in 17 triads.
Its 24 two-valued states assign equal values to a distinguished pair of
distinct rays and hence fail to separate the configuration.  The projector
cones have 230 facets each.  Because the Born table contains genuine
$\mathbb Q(\sqrt2)$ entries, the implemented algebraic branch is numerical and
returns
\begin{equation}
        r_{\Gamma_3}\simeq0.502197722972,
        \label{eq:gamma3-result}
\end{equation}
with maximum numerical dual violation $1.6\times10^{-10}$.  No exact value is
claimed for Eq.~\eqref{eq:gamma3-result}.

\subsection{Tkadlec's nonunital configuration}

Tkadlec's Fig.~2 gives a nonunital Sch\"utte-type configuration in three
dimensions~\cite{tkadlec-96}.  The figure marks 25 rays; the full
suborthoposet contains 37 rays in 26 triads.  The consolidated audit constructs
the full set rather than transcribing it: every orthogonal pair among the 25
marked rays is completed by its unique cross-product ray, and projective
duplicates are removed.  Independent validation then finds 37 rays, 26 triads,
eight two-valued states, and eight rays that are never assigned value one.

The 37-projector cone has 356 facets on each side.  The projector LP
returns
\begin{equation}
        r_{\mathrm{Tkadlec}}\simeq0.547491649817,
        \label{eq:tkadlec-result}
\end{equation}
with maximum equality residual $2.8\times10^{-13}$.  No exact primal--dual
certificate was recovered, so Eq.~\eqref{eq:tkadlec-result} is numerical.

\subsection{Yu--Oh: 13 rays versus 25-ray orthogonal completion}
\label{subsec:yuoh-comparison}

The original Yu--Oh (YO) construction consists of the 13 qutrit rays
\begin{align}
&(1,0,0),(0,1,0),(0,0,1),\nonumber\\
&(0,1,1),(0,1,-1),(1,0,1),(1,0,-1),(1,1,0),(1,-1,0),\nonumber\\
&(1,1,1),(1,1,-1),(1,-1,1),(-1,1,1).
\end{align}
The projector fragment has 24 cone facets.  The simplex linear program gives
\begin{equation}
        r_{\YO13}=\frac38.
        \label{eq:yo13-result}
\end{equation}
In the facet ordering returned by the exact audit program, an optimal certificate can be chosen diagonal, with 18 nonzero diagonal entries and exact residual zero.

The 25-ray version used here is the orthogonal completion of the 13 rays: for every orthogonal pair among the original 13 rays, the cross-product ray completing the triad is added, and duplicates are removed projectively.  The completion contains 25 projectors, its cone has 96 facets, and the linear program returns
\begin{equation}
        r_{\YO25}=\frac{21}{44}\simeq0.47727273.
        \label{eq:yo25-result}
\end{equation}
The larger value is not surprising: the enlarged projector set generates a different operational cone and supplies more Born constraints to be simulated.  The rational audit verifies exact primal and exact dual certificates at $r=21/44$, so this value is an exact optimum for the restricted projector-cone fragment.

The vector-generated closure audit in Table~\ref{tab:closure-results} makes the cost of the restriction explicit.  For the 13-ray set, closing the effects while keeping the original eigenpreparations plus the maximally mixed state gives a numerical optimum near \(9/20\).  Closing the preparation side as well gives a numerical optimum near \(21/44\), equal within the reported numerical precision to the certified value for the 25-projector completion.  Equality of these scalar thresholds is reported as a numerical observation, not as a proof that the underlying cones are identical.

\subsection{Cabello 18--9 and the Peres--Mermin 24--24 completion}

The Cabello--Estebaranz--Garc\'ia-Alcaine 18--9 set is a four-dimensional KS configuration with 18 rays in 9 tetrads~\cite{cabello-96}.  The projector-only fragment is represented in the 10-dimensional real symmetric trace basis.  The cone generated by the 18 projectors has 146 facets.  The numerical optimum rationalizes to $1/3$, and the exact rational audit verifies exact primal and exact dual certificates,
\begin{equation}
        r_{18/9}=\frac13.
        \label{eq:cabello-result}
\end{equation}
so the value is an exact optimum for the restricted projector-cone fragment.

The Peres--Mermin square, analyzed through matrix pencils, yields the 24-24 completion.  The 24 vectors include the Cabello 18-ray set as a subset; the remaining rays supply the orthogonal completion associated with the full 24--24 Peres configuration~\cite{svozil-2024-convert-pra-externalfigures}.  In the projector-only simplex test, the 24 projectors generate a cone with 120 facets.  Writing \(\mathrm{PM24}\) for this Peres--Mermin 24-ray fragment, the numerical optimum rationalizes to $4/9$, and the exact rational audit verifies exact primal and exact dual certificates,
\begin{equation}
        r_{\mathrm{PM24}}=\frac49.
        \label{eq:pm24-result}
\end{equation}
The larger value is consistent with adding projectors and hence strengthening the restricted fragment.  This row is also an exact optimum for the restricted projector-cone fragment.

\begin{table*}[t]
\caption{Projector-cone depolarizing thresholds for the specified labelled prepare-and-measure fragments.  The listed rays are used both as rank-one preparations and rank-one effects, with state-side depolarizing noise $\rho\mapsto \tr(\rho)\id_d/d$.  ``Exact optimum'' means exact primal and dual certificates.  ``Numerical'' means floating-point evidence only.  These values are properties of the stated projector-cone fragments and noise map; they are neither invariants of the underlying hypergraphs nor absolute contextuality strengths.}
\label{tab:results}
\resizebox{\textwidth}{!}{%
\begin{tabular}{lcccccc}
\toprule
Configuration & $d$ & projectors & accessible dim. & facets & $r$ & certificate status\\
\midrule
Firefly logic $L_{12}$ & 3 & 5 & 4 & 5 & $0$ & exact optimum\\
Completed pentagon $C_5$ & 3 & 10 & 6 & 12 & $\approx0.119140568134$ & numerical/algebraic branch\\
Specker bug & 3 & 13 & 6 & 26 & $25/79$ & exact optimum\\
Kochen--Specker $\Gamma_3$ & 3 & 27 & 6 & 230 & $\approx0.502197722972$ & numerical/algebraic branch\\
Tkadlec Fig.~2, nonunital & 3 & 37 & 6 & 356 & $\approx0.547491649817$ & numerical\\
Yu--Oh 13-ray projector fragment & 3 & 13 & 6 & 24 & $3/8$ & exact optimum\\
Yu--Oh 25-ray orthogonal completion & 3 & 25 & 6 & 96 & $21/44$ & exact optimum\\
Cabello--Estebaranz--Garc\'ia-Alcaine 18--9 & 4 & 18 & 10 & 146 & $1/3$ & exact optimum\\
Peres--Mermin 24--24 completion & 4 & 24 & 10 & 120 & $4/9$ & exact optimum\\
\bottomrule
\end{tabular}%
}
\end{table*}

\subsection{Vector-generated operational closures}
\label{subsec:closure-results}

Table~\ref{tab:closure-results} reports the second audit, in which the same
vector sets are closed under the operational data generated by their
orthogonality relations.  The \emph{closed-effects} rows use the original
rank-one eigenpreparations together with the maximally mixed state, and close
the effect side under the unit, residual sharp-measurement outcomes,
complements, coarse grainings, and equality of operators.  The \emph{closed
effects+preparations} rows also add normalized versions of the closed effects
as preparations.  The table does not include the algebraic pentagon, whose
exact closure would require an algebraic-number polyhedral backend.

The closed-effects-plus-preparations variant is an additional modelling choice,
not a consequence of the original ray hypergraph: it assumes that every
normalized closed measurement event is also available as a preparation.  It
therefore defines a stronger operational fragment than the closed-effects
variant.

\begin{table*}[t]
\caption{Vector-generated operational-closure thresholds.  Here
\((|\Omega|,|E|)\) gives the number of preparation and effect generators after
operator equality has been quotiented, and \((f_\Omega,f_E)\) gives the numbers
of state- and effect-cone facets.  The entries are numerical outputs of the
closure script; every entry in the \(r\)-column is numerical, and displayed
fractions are rationalizations of floating-point optima rather than exact
values.  No row in this table is claimed as an exact-certified optimum.}
\label{tab:closure-results}
\resizebox{\textwidth}{!}{%
\begin{tabular}{llcccccl}
\toprule
Configuration & closure variant & $d$ & $(|\Omega|,|E|)$ & $(f_\Omega,f_E)$ & solver & $r_{\rm num}$ & status\\
\midrule
Firefly logic $L_{12}$ & closed effects & 3 & $(6,11)$ & $(5,5)$ & primal & $\approx0$ & numerical\\
Firefly logic $L_{12}$ & closed effects+preparations & 3 & $(11,11)$ & $(5,5)$ & primal & $\approx0$ & numerical\\
Yu--Oh 13 & closed effects & 3 & $(14,51)$ & $(24,96)$ & primal & $\approx9/20$ & rationalized\\
Yu--Oh 13 & closed effects+preparations & 3 & $(51,51)$ & $(96,96)$ & primal & $\approx21/44$ & rationalized\\
Yu--Oh 25 completion & closed effects & 3 & $(26,51)$ & $(96,96)$ & primal & $\approx21/44$ & rationalized\\
Yu--Oh 25 completion & closed effects+preparations & 3 & $(51,51)$ & $(96,96)$ & primal & $\approx21/44$ & rationalized\\
Specker bug & closed effects & 3 & $(14,27)$ & $(26,26)$ & primal & $\approx25/79$ & rationalized\\
Specker bug & closed effects+preparations & 3 & $(27,27)$ & $(26,26)$ & primal & $\approx25/79$ & rationalized\\
Kochen--Specker $\Gamma_3$ & closed effects & 3 & $(28,55)$ & $(230,230)$ & cutting dual & $\approx0.502197722951$ & numerical\\
Kochen--Specker $\Gamma_3$ & closed effects+preparations & 3 & $(55,55)$ & $(230,230)$ & cutting dual & $\approx0.502197722951$ & numerical\\
Tkadlec Fig.~2, nonunital & closed effects & 3 & $(38,75)$ & $(356,356)$ & cutting dual & $\approx0.547491649818$ & numerical\\
Tkadlec Fig.~2, nonunital & closed effects+preparations & 3 & $(75,75)$ & $(356,356)$ & cutting dual & $\approx0.547491649818$ & numerical\\
Cabello 18--9 & closed effects & 4 & $(19,121)$ & $(146,120)$ & primal & $\approx4/9$ & rationalized\\
Cabello 18--9 & closed effects+preparations & 4 & $(121,121)$ & $(120,120)$ & primal & $\approx4/9$ & rationalized\\
Peres--Mermin/Peres 24 & closed effects & 4 & $(25,139)$ & $(120,120)$ & primal & $\approx4/9$ & rationalized\\
Peres--Mermin/Peres 24 & closed effects+preparations & 4 & $(139,139)$ & $(120,120)$ & primal & $\approx4/9$ & rationalized\\
\bottomrule
\end{tabular}%
}
\end{table*}

The closure rows sharpen the comparison with the projector-cone proxy.  Firefly
remains simplex-embeddable after closure for this specified vector-generated
fragment.  For Yu--Oh 13 the numerical optimum rationalizes from the
exact-certified projector value \(3/8\) to \(9/20\) when the effect closure is
added, and to \(21/44\) when the preparation side is closed as well.  The
25-ray Yu--Oh completion is unchanged to the reported numerical precision under
this closure.  Both Specker-bug closures return approximately $25/79$, equal
within numerical precision to the exact projector-cone optimum.  Both
$\Gamma_3$ closures return approximately $0.502197722951$; their cutting-dual
maximum violations are below $7\times10^{-15}$.  For the Tkadlec nonunital
configuration, both closure variants return approximately
\(0.547491649818\), equal within numerical precision to the projector-cone
value; the corresponding cutting-dual maximum violations are
\(7.1\times10^{-11}\) and \(5.9\times10^{-12}\), respectively.
Cabello 18--9 has exact-certified projector-cone value \(1/3\); after effect
closure the numerical optimum rationalizes to \(4/9\), numerically consistent with the
Peres--Mermin 24-ray completion.  Peres--Mermin/Peres 24 is unchanged to the
reported numerical precision at \(4/9\) under the reported closures.

\section{Interpretation of the tables}

Table~\ref{tab:results} answers one narrow question: how much state-side depolarizing noise is needed before the Born bilinear form on the cones generated by the listed projectors admits a simplex embedding?  It does not give chromatic numbers, counts of two-valued states, or full operational robustnesses for complete contextuality scenarios.  Table~\ref{tab:closure-results} answers a second, more operational question for the source-derived vector rows: what changes when the unit, residual sharp-measurement outcomes, complements, coarse grainings, and simple preparation closures generated by the same vectors are added?

Rows marked ``exact optimum'' are certified by rational primal and dual witnesses.  Rows marked ``numerical'' or ``numerical/algebraic branch'' remain floating-point evidence rather than independent exact certificates.

The comparisons should therefore be read as diagnostics rather than rankings.  $L_{12}$ is a classically representable partition logic and has $r=0$ even after closure.  The completed pentagon gives a small positive numerical value, as expected for the first cyclic probability-level example.  The Specker bug, $\Gamma_3$, and the Tkadlec nonunital configuration exhibit nonfull, nonseparating, and nonunital valuation defects, respectively.  The Yu--Oh and Cabello/Peres--Mermin pairs show that orthogonal or operational completion can change the cone and can raise the threshold.  The useful content is the uniform computation together with the certificate status of each row.

\section{Discussion}
\label{sec:discussion}

The examples support a modest conclusion.  Valuation tests, simplex-embedding tests, and product-rule tests often agree on canonical KS configurations, but they need not expose the same obstruction after the data have been restricted.

This is clearest for operator-valued parity proofs.  If an argument is fully resolved into spectral projectors, it may become an ordinary projective quantum-logic problem, as in the Peres--Mermin transcription to the 24-ray completion.  If the same argument is kept at the level of coarse-grained commuting observables, the relevant extra assumption may instead be multiplicativity of assigned values.  GHZ isolates that point: the collective joint eigensystem is one Boolean block, but the attempted decomposition of its global product observables into context-independent local factors gives the wrong product sign.

The computational tables have the same limited status.  Table~\ref{tab:results} compares selected projector cones and reports which values are exactly certified.  Table~\ref{tab:closure-results} compares a specified vector-generated operational closure of the source-derived rows.  It is not a universal ranking of contextuality.  The Specker bug, $\Gamma_3$, and the Tkadlec configuration realize nonfullness, nonseparation, and nonunitality of their two-valued-state spaces; Yu--Oh 13 and Yu--Oh 25, and likewise Cabello 18--9 and Peres--Mermin/Peres 24, generate different cones after orthogonal or operational completion.

Partition logics and chromatic colorability sit naturally on the incidence side of this comparison.  Partition logics show that complementarity and non-Boolean pasting do not by themselves force nonclassical probabilities.  Strong colorability asks for a global decomposition of the unit assignment into spectral labels, a stricter condition than the mere existence of two-valued states.  Both diagnostics are useful, but neither is the same object as a simplex robustness computed from a probability table.

\section{Conclusion}

The useful distinction is therefore not between three new kinds of physics, but between three choices of data: incidence data for valuation, partition-logic, and coloring questions; convex-operational data for simplex-embedding questions; and operator data for functional-composition and product-rule questions.

The resulting distinctions are concrete.  A partition-logically representable
hypergraph can nevertheless support nonclassical Born probabilities in a
quantum realization.  Conversely, a closed single Boolean context is
simplex-classical.  The GHZ contradiction is therefore not located in the
collective joint spectral context itself, but in the additional
product-preserving identification of its global outcomes with pre-existing
local Pauli values.  Likewise, simplex nonembeddability, KS uncolorability,
and chromatic contextuality diagnose different possible failures of
classicality; none is merely a reformulation of the others.

With that restriction in view, the two main contributions are limited but concrete: the GHZ discussion isolates multiplicative product preservation as an additional assumption not contained in the Boolean algebra of the collective joint spectral measurement, and the simplex audits give a uniform set of depolarizing thresholds for both projector-cone fragments and vector-generated closures.  The projector-cone rows explicitly marked \emph{exact optimum} in Table~\ref{tab:results} have exact rational primal and dual certificates.  The closure rows in Table~\ref{tab:closure-results} show how much the reported thresholds can change once unit effects, residual sharp outcomes, complements, coarse grainings, and normalized event preparations are restored, but they remain numerical unless exact certificates are supplied.

Most importantly, no number in either table is a contextuality strength of a
named hypergraph in isolation.  Each number belongs to a particular labelled
preparation--effect fragment, its specified operational closure (if any), and
the chosen state-side depolarizing map.  The tables are therefore comparative
audits of those fragments, while the incidence, simplex, and product-rule tests
remain distinct diagnostics of different retained structures.

\section*{Data and code availability}

The supplementary archive, available at \url{https://svozil.github.io/publications/2026-simplex-simplex_supplement_bundle.zip}, contains one self-contained program,
\texttt{simplex\_audit\_rebuilt.py}, the complete \texttt{audit\_v3} output
directory, the explicit vector-and-context catalogue, the supplemental text,
and SHA-256 checksums.  The archived program hash is
\nolinkurl{d2298aeaab2d47a42e94e9077ece208df8042ddf6603d6fd0252d8fc184e1596};
the same hash is recorded by the run manifest.  Exact regeneration of rational
facets and certificates requires a fraction/GMP-enabled
\texttt{pycddlib}/cddlib installation.  The default command
\texttt{python simplex\_audit\_rebuilt.py --outdir audit\_v3} validates every
source configuration, runs every projector and closure task, and reuses a task
only when its input fingerprint matches.  Software versions are recorded in
the \texttt{runtime} object of \texttt{audit\_v3/audit\_report.json}.

\begin{acknowledgments}
This research was funded in whole or in part by the \textit{Austrian Science Fund (FWF)} [Grant \href{https://doi.org/10.55776/PIN5424624}{digital object identifier (DOI): 10.55776/PIN5424624}].
The author acknowledges TU Wien Bibliothek for financial support through its Open Access Funding Programme.

OpenAI Codex (GPT-5.6) was used to assist with manuscript criticism, literature
organization, checking derivations, and editorial revision.  The author
directed its use, independently verified the mathematical arguments and cited
sources, revised the resulting text, and assumes full responsibility for the
manuscript.
\end{acknowledgments}

\appendix

\section{Computational audit data}
\label{app:audit}

This appendix records the computational status behind Tables~\ref{tab:results} and~\ref{tab:closure-results}.  It is included because the distinction between numerical optima, exact primal certificates, and exact dual certificates is essential for interpreting the tables.

The exact audit uses the accessible-fragment form of Eq.~\eqref{eq:stage3-audit}.  Integer or rational rays are converted to raw real-symmetric projector coordinates.  Effects are stored in dual coordinates, \(\widetilde I_E=GI_E\), where $G=\operatorname{diag}(\underbrace{1,\ldots,1}_{d},\underbrace{2,\ldots,2}_{d(d-1)/2})$, and cddlib is used over $\mathbb Q$ to obtain rational facet matrices whenever possible.  A numerical linear program first proposes $r$ and $\sigma$.  The proposed rational values are then checked exactly by verifying
\begin{equation}
        H_E^T\sigma H_\Omega
        =\widetilde I_E^T\bigl[(1-r)\operatorname{Id}_{\mathcal A}+rD\bigr]I_\Omega,
        \qquad \sigma\ge_{\mathrm e}0.
        \label{eq:exact-primal-check}
\end{equation}
This is a primal feasibility certificate.  Exact optimality additionally requires a rational dual witness.  The audit verifies such dual witnesses for all rational rows reported as exact optima in Table~\ref{tab:results}.

For the Specker bug, the ray representatives lie in $\mathbb Q(\sqrt2)$ but
all projector overlaps are rational.  The program rank-factorizes that Born
table over $\mathbb Q$ and applies the same cone-facet and primal--dual checks
in the resulting six-dimensional operational coordinates.  For $\Gamma_3$,
some Born overlaps remain in $\mathbb Q(\sqrt2)\setminus\mathbb Q$; its
operational rank factorization, hull facets, and LP are therefore numerical.

The dual certificate is as follows.  Put
\begin{equation}
        M_0=\widetilde I_E^TI_\Omega,\qquad
        M_1=\widetilde I_E^T(D-\operatorname{Id}_{\mathcal A})I_\Omega,
\end{equation}
and let \(C_{\ell k}=(h^E_\ell)^T h^\Omega_k\) denote the rank-one coefficient
matrix multiplying \(\sigma_{\ell k}\).  Omitting the inactive upper bound
\(r\le1\) gives the lower-bounded primal LP
\begin{equation}
        \begin{aligned}
        \min\ & r\\
        \text{s.t.}\quad&
        \sum_{\ell,k}\sigma_{\ell k}C_{\ell k}-rM_1=M_0,\\
        &\sigma_{\ell k}\ge0,\qquad r\ge0 .
        \end{aligned}
        \label{eq:certificate-primal}
\end{equation}
All reported certified optima satisfy \(0\le r\le1\), so a certificate for
Eq.~\eqref{eq:certificate-primal}, together with primal feasibility for the
bounded problem, certifies the implemented LP.  With the Frobenius inner product
\(\langle A,B\rangle=\tr(A^TB)\) on the accessible coefficient matrices, the dual is
\begin{equation}
        \begin{aligned}
        \max\ & \langle Y,M_0\rangle\\
        \text{s.t.}\quad&
        \langle Y,C_{\ell k}\rangle\le0
        \quad\text{for all }\ell,k,\\
        &-\langle Y,M_1\rangle\le1 .
        \end{aligned}
        \label{eq:certificate-dual}
\end{equation}
A rational matrix \(Y\) satisfying these inequalities and
\(\langle Y,M_0\rangle=r_*\) proves that no feasible point has \(r<r_*\).
Together with a rational nonnegative \(\sigma\) satisfying
Eq.~\eqref{eq:exact-primal-check} at \(r_*\), this is the primal--dual
certificate used for the ``exact optimum'' rows.

The projector-cone Python audit returned the status summarized in Table~\ref{tab:appendix-audit}.  The table is intentionally diagnostic rather than a second robustness table: its purpose is to show which entries are exact optima and which remain numerical.  For each rational row marked ``exact optimum,'' exact primal feasibility is checked by Eq.~\eqref{eq:exact-primal-check}, and exact optimality is certified by a rational dual witness recovered from the numerical dual active set when direct rationalization is insufficient.

\begin{table*}[t]
\caption{Audit status of the simplex computations.  Here $k_\Omega$ and $k_E$ are the accessible state and effect dimensions after Stage-1 reduction.  The rational rows marked ``exact optimum'' have both exact primal and exact dual certificates.}
\label{tab:appendix-audit}
\resizebox{\textwidth}{!}{%
\begin{tabular}{lcccccccl}
\toprule
Configuration & projectors & $k_\Omega$ & $k_E$ & facets & $r$ & exact primal & exact dual & note\\
\midrule
Firefly logic $L_{12}$ & 5 & 4 & 4 & 5 & $0$ & yes & yes & exact optimum\\
Yu--Oh 13 & 13 & 6 & 6 & 24 & $3/8$ & yes & yes & exact optimum\\
Yu--Oh 25 completion & 25 & 6 & 6 & 96 & $21/44$ & yes & yes & exact optimum\\
Specker bug & 13 & 6 & 6 & 26 & $25/79$ & yes & yes & exact optimum\\
Kochen--Specker $\Gamma_3$ & 27 & 6 & 6 & 230 & $\approx0.502197722972$ & no & no & algebraic/numerical\\
Tkadlec Fig.~2, nonunital & 37 & 6 & 6 & 356 & $\approx0.547491649817$ & no & no & numerical\\
Cabello 18--9 & 18 & 10 & 10 & 146 & $1/3$ & yes & yes & exact optimum\\
Peres--Mermin/Peres 24 & 24 & 10 & 10 & 120 & $4/9$ & yes & yes & exact optimum\\
Completed pentagon $C_5$ & 10 & 6 & 6 & 12 & $\approx0.119140568134$ & no & no & algebraic/numerical branch\\
\bottomrule
\end{tabular}%
}
\end{table*}

The completed pentagon is not exact-certified by the rational backend because the coordinates in Eq.~\eqref{eq:kcbs-rays} are algebraic.  The numerical branch found 10 projectors, full accessible dimension six, 12 facets, numerical robustness $r=0.119140568134$, maximum residual approximately $1.8\times10^{-13}$, and 16 active $\sigma$ entries.  The $\Gamma_3$ branch found 27 projectors, 230 facets per cone, and $r=0.502197722972$, with maximum dual violation approximately $1.6\times10^{-10}$.  The rational Tkadlec nonunital input has exact facets, but its optimum was not recovered as an exact primal--dual pair; its displayed value is therefore numerical.  Exact certification of the pentagon and $\Gamma_3$ would require an algebraic-number polyhedral backend or a rationally equivalent reformulation.

The same program performed the vector-closure audit.  For each source-derived vector
set it generated residual sharp-measurement effects, complements, coarse
grainings, and duplicate-operator quotients from the vector data itself.  The
run reported in Table~\ref{tab:closure-results} used exact cddlib facets
throughout the rational closure rows.  The large $\Gamma_3$ and Tkadlec rows were solved by a
cutting-dual linear program; this is numerically much smaller than the primal
\(\sigma\)-program but does not return an exact primal certificate.
Consequently every value in Table~\ref{tab:closure-results} is reported as
numerical.  The complete command
\texttt{python simplex\_audit\_rebuilt.py --outdir audit\_v3}
finished without error and wrote the archived manifests.

\bibliography{svozil}

\end{document}